\documentclass[twocolumn]{aastex701}
\usepackage{newtxtext,newtxmath}  
\usepackage{amsmath}
\usepackage{graphicx}
\usepackage{float}
\usepackage{placeins}

\newcommand{\equ}[1]{eq.~(\ref{eq:#1})}

\newcommand{\equnp}[1]{eq.~\ref{eq:#1}}
\newcommand{\se}[1]{\S\ref{sec:#1}}

\newcommand{\Fig}[1]{Figure~\ref{fig:#1}}

\newcommand{\be}{\begin{equation}}
\newcommand{\ee}{\end{equation}}
\newcommand{\ba}{\begin{align}}
\newcommand{\ea}{\end{align}}
\newcommand{\bad}{\begin{equation} \begin{aligned}}
\newcommand{\ead}{\end{aligned} \end{equation}}
\newcommand{\bea}{\begin{eqnarray}}
\newcommand{\eea}{\end{eqnarray}}
\def\ra{\rangle}
\def\la{\langle}

\newcommand{\bul}{$\bullet\ $}

\newcommand{\msun}{M_\odot}
\newcommand{\Msun}{M_\odot}

\newcommand{\Zsun}{Z_\odot}

\newcommand{\ifm}[1]{\relax\ifmmode#1\else$\mathsurround=0pt #1$\fi}
\newcommand{\kms}{\ifmmode\,{\rm km}\,{\rm s}^{-1}\else km$\,$s$^{-1}$\fi}

\newcommand{\Mpc}{\,{\rm Mpc}}
\newcommand{\kpc}{\,{\rm kpc}}
\newcommand{\pc}{\,{\rm pc}}

\newcommand{\Gyr}{\,{\rm Gyr}}

\newcommand{\Myr}{\,{\rm Myr}}

\newcommand{\cmc}{\,{\rm cm}^{-3}}

\newcommand{\ltsima}{$\; \buildrel < \over \sim \;$}
\newcommand{\lsim}{\lower.5ex\hbox{\ltsima}}
\newcommand{\gtsima}{$\; \buildrel > \over \sim \;$}
\newcommand{\gsim}{\lower.5ex\hbox{\gtsima}}
\newcommand{\prop}{\propto}
\newcommand{\dd}{\rm d}

\def\Mv{M_{\rm vir}} % 

\def\Mveight{M_{{\rm v},10.8}}

\def\Rv{R_{\rm v}}
\def\Vv{V_{\rm v}}

\def\Ms{M_{*}}

\def\Sig1{\Sigma_1}

\def\Md{M_{\rm d}}
\def\Rd{R_{\rm d}}
\def\Hd{H_{\rm d}}

\def\Mc{M_{\rm c}}
\def\Rc{R_{\rm c}}

\def\Vc{V_{\rm c}}

\def\td{t_{\rm dyn}}

\def\tseg{t_{\rm seg}}
\def\tdf{t_{\rm df}}

\def\eps2{\epsilon_{-2}}

\def\eps{\epsilon}

\def\ssim{\!\sim\!}
\def\seq{\!=\!}
\def\ssimeq{\!\simeq\!}
\def\sequiv{\!\equiv\!}
\def\sgt{\!>\!}
\def\slt{\!<\!}
\def\sgsim{\!\gsim\!}
\def\slsim{\!\lsim\!}
\def\sgeq{\!\geq\!}
\def\sleq{\!\leq\!}

\def\sll{\!\ll\!}
\def\sdash{\!-\!}
\def\stimes{\!\times\!}
\def\sprop{\!\propto\!}

\def\Rc{R_{\rm c}}

\def\Mcsix{M_{{\rm c},6}}

\def\Mcsix{M_{{\rm c},6}}

\def\z110{(1+z)_{10}}
\def\lam25{\lambda_{.025}}
\def\Sq067{\Sigma_{Q=0.67}}
\def\Sigq067{\Sigma_{Q=0.67}}

\def\Ms{M_{*}}

\def\x{{x}}
\def\y{\gamma}
\def\Mc{M_{\rm c}}
\def\Rc{R_{\rm c}}
\def\Vc{V_{\rm c}}
\def\Ec{E_{\rm c}}
\def\Mp{M_{\rm p}}
\def\Vp{V_{\rm p}}
\def\np{n_{\rm p}}
\def\Np{N_{\rm p}}
\def\Rd{R_{\rm d}}
\def\Hd{H_{\rm d}}

\def\Vesc{V_{\rm esc}}
\def\zffb{z_{\rm ffb}}
\def\Fd{F_{\rm d}}
\def\Fh{F_{\rm h}}
\def\Mcsix{M_{\rm c,6}}
\def\Rcseven{R_{\rm c,7}}

\defcitealias{dekel23}{D23}
\defcitealias{li23}{L23}

\begin{document}

\title[From FFB to LRD via Compaction]
{From Feedback-Free Star Clusters to Little Red Dots via Compaction}

\author[orcid=0000-0003-4174-0374]{Avishai Dekel}
\affiliation{Racah Institute of Physics, The Hebrew University, Jerusalem 91904, Israel}
\email{dekel@huji.ac.il}  

\author[orcid=0000-0003-0250-3827]{Dhruba Dutta Chowdhury} 
\affiliation{Racah Institute of Physics, The Hebrew University, Jerusalem 91904, Israel}
\email[show]{dhruba.duttachowdhury@mail.huji.ac.il}

\author[orcid=0009-0004-0973-3232]{Sharon Lapiner}
\affiliation{Racah Institute of Physics, The Hebrew University, Jerusalem 91904, Israel}
\email{sharon.lapiner@mail.huji.ac.il}

\author[orcid=0000-0003-1604-1272]{Zhiyuan Yao}
\affiliation{Racah Institute of Physics, The Hebrew University, Jerusalem 91904, Israel}
\email{zhiyuan.yao@mail.huji.ac.il}
 
\author[orcid=0000-0002-6462-6657]{Shmuel Gilbaum}
\affiliation{Racah Institute of Physics, The Hebrew University, Jerusalem 91904, Israel}
\email{shmuel.gilbaum@mail.huji.ac.il}

\author[orcid=0000-0002-8680-248X]{Daniel Ceverino}
\affiliation{Departamento de Fisica Teorica and CIAFF, Facultad de Ciencias, Universidad
Autonoma de Madrid, Cantoblanco, 28099 Madrid, Spain}
\email{daniel.ceverino@uam.es}

\author[orcid=0000-0001-5091-5098]{Joel Primack}
\affiliation{Department of Physics and SCIPP, University of California, Santa Cruz, CA
96064, USA}
\email{joel@ucsc.edu}

\author[orcid=0000-0002-6748-6821]{Rachel Somerville}
\affiliation{Center for Computational Astrophysics, Flatiron Institute, 162 5th Avenue,
New York, NY 10010, USA}
\email{rsomerville@flatironinstitute.org}

\author[orcid=0000-0001-7689-0933]{Romain Teyssier}
\affiliation{Department of Astrophysical Sciences, Princeton University, 4 Ivy Lane,
Princeton, NJ 08540, USA}
\email{teyssier@princeton.edu}

\begin{abstract}
We address the origin of the Little Red Dots (LRDs) seen by JWST at cosmic morning ($z \seq 4 \sdash 8$) as compact stellar systems with over-massive black holes (BHs). We propose that LRDs form naturally after feedback-free starbursts (FFB) in thousands of star clusters and following wet compaction. Analytically, we show how the clusters enable efficient dry migration of stars and BHs to the galaxy center by two-body segregation and dynamical friction against the disk. The clusters merge to form compact central stellar systems as observed. Mutual tidal stripping does not qualitatively affect the analysis. The young, rotating clusters are natural sites for the formation of BH seeds via rapid core collapse. The migrating clusters carry the BH seeds, which merge into central super-massive BHs (SMBHs). Compactions are required to deepen the potential wells such that the SMBHs are retained after post-merger gravitational-wave recoils, locked to the galaxy centers. Using cosmological simulations at different epochs, with different codes and physical recipes, we evaluate the additional growth of LRD-matching compact central stellar systems by global compaction events. Adding to the dry growth by cluster mergers, the compactions can increase the escape velocities to retain the SMBHs. The LRDs appear at $z \ssim 8$, after the formation of FFB clusters, and disappear after $z \ssim 4$ when the stellar mass is above $10^9\msun$ by growing post-compaction blue disks around the nuclear LRDs. The LRD abundance is expected to be $\sim\! 10^{-5} \sdash 10^{-4}\Mpc^{-3}$, increasing from $z \ssim 4$ to $z\ssim 8$.
\end{abstract}

\keywords{\uat{Galaxies}{594} -- \uat{Galaxies}{595} -- \uat{Galaxies}{600} -- \uat{Galaxies}{734}}

%%%%%%%%%%%%%%%%%%%%%%%%%%%%%% 1
\section{Introduction}
\label{eq:intro}

%==========================
\subsection{Little Red Dots}

%observed LRD: compact and BH
JWST observations reveal the existence of a distinct population of galaxies termed Little Red Dots (LRDs), detected at cosmic morning in the redshift range $z \seq 4-8$ \citep[e.g.,][]{barro24a,barro24b,baggen24,pacucci24a,pacucci24b,casey24,casey25,greene25,akins25a,akins25b,akins25c,pacucci25a,pacucci25b,taylor25a,taylor25b}. Their high abundance is roughly $1.5\sdash 2$ dex below the total abundance of galaxies at the same UV magnitude \citep[e.g.,][]{kokorev24, greene24,kocevski25}. The LRDs are characterized by compact, mostly unresolved red optical sizes. They tend to show a so called ``V-shaped" spectral energy distribution (SED), blue in the UV and red in optical light. They show broad permitted lines, with a flat IR spectrum, no far-IR/sub-mm emission and weak near-to-mid IR emission, and some show Balmer breaks. Typically, no X-ray or radio waves are detected.

\smallskip
The two extreme interpretations of LRDs are that they are either dominated by black holes (BHs) or by stars \citep[e.g.,][]{pacucci24a,pacucci24b}. The BH-dominated interpretation would require a very high ratio of BH-to-stellar mass compared to low redshifts and an explanation for the undetected X-rays. The stellar-dominated interpretation would require extremely high stellar densities, not seen in other galaxies.   

\smallskip
A hybrid interpretation is plausible, where the LRDs consist of both compact stellar systems and over-massive black holes. The hybrid model is supported by the observational finding of \citet{jones25} that most of the galaxies with over-massive BHs are LRDs. They examined a sample of 70 broad-line Active Galactic Nuclei (AGN) identified at $ z \seq 3 \sdash 7$ using NIRSpec/G395M spectroscopy from the CEERS, JADES, and RUBIES JWST surveys. As can be seen in their illuminating Fig.~6, they find that $43\%$ of the sample are classified as LRDs, a fraction that increases with redshift. These LRDs with BHs tend to be of lower stellar masses than the non-LRD AGN, with the BH-to-stellar mass ratio 1-2 dex above the low-redshift AGN and increasing with redshift. They have BH masses in the range $10^{6.7} \sdash 10^{8.7}\msun$, high BH-to-stellar mass ratios of $0.01 \sdash 0.1$, 
and estimated stellar masses of $10^8 \sdash 10^9\msun$.

\smallskip
Before addressing the detailed properties of LRDs, the main basic challenge addressed here is to explain the formation of compact stellar systems with over-massive black holes at cosmic morning, and their apparent  disappearance at later times.

%==========================
\subsection{FFB clusters}

%FFB clusters 
A straightforward physical model predicts a high star-formation efficiency, at the level of $0.2 \sdash 0.3$ for the global ratio of stars to accreted gas, under the conditions of high density and moderate metallicity that are valid in massive galaxies at cosmic dawn \citep{dekel23,li24}. It is due to feedback-free starbursts (FFB) in thousands of star clusters
within each galaxy. This is as opposed to the a star formation efficiency lower than $0.1$ at lower redshifts, where feedback from stars and from AGN is expected to suppress star formation. The FFB phase is expected to occur at $z \sgt 8$, in dark-matter halos of mass $\Mv \sgt 10^{10.5}\msun$, with the threshold mass dropping steeply as a function of redshift, $\Mv \sprop\! (1+z)^{-6.2}$. The typical resultant galaxy is a rather compact rotating disk of half-mass radius $\sim\! 300\pc$, consisting of thousands of rotating star clusters of masses in the range  $10^4 \sdash 10^7\msun$ and radii of order $10\pc$. These galaxies are predicted to have low gas fraction ($f_{\rm gas} \ssim 0.01 \sdash 0.1$), relatively low metallicity ($Z \ssim 0.01 \sdash 0.1\Zsun$), and low dust attenuation ($A_{\rm UV} \ssim 0.5$). We propose that the FFB clusters serve as a key element in the formation of LRDs, both in the origin of a compact stellar system and in the growth of an over-massive central black hole.

\smallskip % Chen
The first cosmological simulation of a galaxy at cosmic dawn that is marginally capable of capturing the FFB main physical processes in clusters indeed reveals the formation of thousands of clusters \citep[][]{chen26}. This is a simulation utilizing the GIZMO code \citep{hopkins15}, with $\sleq 3\pc$ gas resolution, which incorporates a $3.4\Myr$ delay in supernova feedback. The simulated cluster properties are found to largely resemble the FFB model predictions, including total mass, mass function, high densities, bursty star-formation history, high star-formation efficiency, low gas fraction, low metallicity, weak dust obscuration, and the migration of the massive clusters towards the galaxy center.

\smallskip % clusters observed
Cluster-dominated galaxies are indeed observed at cosmic morning. JWST observations using gravitational lensing reveal galaxies in the redshift range $z \seq 4 \sdash 10$ in which the stellar mass is dominated by young massive star clusters \citep{claeyssens23,vanzella23,fujimoto23,messa24,adamo24,mowla24a,mowla24b}. They are descriptively dubbed ``Sunrise Arc", ``Cosmic Grapes", ``Firefly Sparkle", ``Cosmic Gems", and so on. These galaxies typically show five to ten clusters each, with half-mass radii of $\sim\! 10\pc$ ($1 \sdash 100\pc$), stellar masses of $\sim\! 10^{6.5} \msun$ ($10^5 \sdash 10^9\msun$), surface densities of $\sim\! 10^4 \msun \pc^{-2}$ ($10^3 \sdash 10^5\msun\pc^{-2}$), stellar ages of $\sim\! 10\Myr$ ($1 \sdash 35\Myr$), and relatively low metallicities of $<\! 0.1\Zsun$. They thus resemble in may ways the massive end of clusters predicted in the FFB scenario. The estimated overall stellar masses of the host galaxies are sometimes smaller than predicted by the original FFB model, but it is possible that the less massive clusters and the background stellar mass are not properly resolved and are therefore underestimated.

%==============================
\subsection{Black Hole Growth}

%BH growth 
Another central issue in the origin of LRDs is the rapid growth of super-massive central black holes (SMBHs). A feasibility study \citep{dekel25_bh} indicates that the FFB scenario may provide a natural setting for such a black hole growth. The scenario involves the formation of seed BHs, of $\sim 10^4\msun$, by sped-up core collapse in the young, rotating FFB star-clusters \citep{spitzer71,hachisu79,portegies04,devecchi09,katz15,rantala24}. It is followed by migration of the clusters and BHs to the galactic-disk centers, driven by dynamical friction and two-body segregation that act on the clusters and the BHs. For the BH seeds to merge into SMBHs that are retained in the galaxy centers, the merger remnants have to overcome the bottleneck introduced by gravitational-wave (GW) recoils \citep{pretorius05,campanelli06,baker06}. The ejection velocities in the case of BHs with comparable masses and non-negligible misaligned spins, as expected in galactic disks that are not unrealistically cold, are typically on the order of several hundred $\kms$. These are larger than the escape velocities of $\sim\! 200\kms$ expected in the typical cosmic-dawn galaxies produced in the FFB phase.

\smallskip %BH growth needs compaction
The lesson from the analysis of \citet{dekel25_bh} is that a necessary condition for the growth of the central SMBH is a deepening of the central galactic potential well. This turns out to be a natural result of wet compaction events (see below). Beyond overcoming the GW recoils, such a deepening of the potential well would prevent suppression of the dynamical-friction-driven inspiral into the galactic center by core stalling \citep{read06,goerdt06,kaur18,dutta19,banik21}. It could also assist in overcoming the final parsec problem \citep{begelman80} by providing central gas. Finally, it would stop the off-center wandering of the SMBH, locking it to the galaxy center \citep{lapiner21}.

%==========================
\subsection{Wet Compaction Events}

%Compaction
Both cosmological simulations \citep[e.g.,][]{zolotov15,lapiner23} and observations \citep[e.g.,][]{barro13,dokkum15,barro17,degraaff24} reveal that most galaxies undergo throughout their histories gas-rich (``wet") compaction events into compact starbursts (termed ``blue nuggets"). These events induce major transitions in the galaxy properties, and in particular trigger quenching of star formation into ``red nuggets". As discussed in \citet{lapiner23}, the compaction processes are due to drastic angular-momentum losses. About half of these are caused by mergers and the rest by colliding counter-rotating streams, recycling fountains from the galactic disks, disk-instability-driven radial transport \citep{db14} and possibly other mechanisms. 

\smallskip % golden mass
The major compaction events, those that trigger a decisive long-term quenching process and a transition from central dark-matter to baryon dominance, tend to occur near the golden mass, halos of masses $10^{11} \sdash 10^{12}\msun$ \citep{dekel19_gold} at all redshifts. This is seen in cosmological simulations \citep{tomassetti16} and in machine-learning-aided comparisons to observations \citep{huertas18}. This preferred mass scale for major compaction events seems to be due, at least partly, to the physical processes of supernova feedback and hot circum-galactic medium (CGM), which tend to suppress compaction attempts at lower and higher masses, respectively. At cosmic dawn, the FFB halos are indeed of masses $\sim\!10^{11}\msun$, in the vicinity of the golden mass, and they correspond to high-sigma peaks of $>\!4\!\sigma$ \citep{dekel25_post}, which tend to have more frequent and stronger compaction events due to multiple colliding streams with low angular momentum \citep{dubois12}.

\smallskip % compaction drives quenching and disk
The compaction events trigger a sequence of morphological changes, which may be relevant both to the beginning and the end of the LRD phase. The compaction events drive central starbursts in the form of blue nuggets, followed by inside-out quenching to quiescent red nuggets, as a result of gas depletion due to star formation and the associated outflows \citep{tacchella16_ms, tacchella16_prof}.

\smallskip % sims
We use here cosmological simulations, utilizing a variety of codes and physical models at different epochs, to quantify the compaction-driven growth of compact, massive central stellar systems, beyond their growth by dry mergers between the clusters. These compact central stellar systems could plausibly be identified with the stellar systems of LRDs, and could possibly increase the escape velocities to the level that is required for retaining the SMBHs at the centers against GW recoils.

%===============================
\subsection{Disappearance of LRDs}

% disk/ring
Subsequently, the compaction events lead to the formation of extended gas disks and rings, fueled by new incoming cold streams \citep{tacchella16_prof, dekel20_ring}. The compact massive nuggets (or bulges) stabilize the disks/rings against gravitational instability. This suppresses star formation in the outskirts \citep[``morphological quenching",][]{martig09} and limits the rate of instability-driven radial gas transport toward the center, maintaining long-lived, star forming disks/rings \citep{dekel20_ring,dutta24}. These disks/rings tend to appear above the golden mass.

\smallskip % post-compaction end of LRD
Such a disk would mark the end of a naked red nugget, or LRD, by adding a blue extended envelope. Indeed, using JWST/CEERS observations, \citet{billand25} identify a population of galaxies at $z \slt 4$ that could represent the descendants of LRDs after developing blue peripheries of recently formed young stars around their red cores. The consistency is in terms of the properties of the red core and the number densities of the populations. It turns out that long-lived extended disks are coincidentally expected to be favored above
the same golden mass of $\Mv \ssim 10^{11}\msun$ due to the fact that merger-driven disk flips become less frequent than the disk orbital time above this mass at all redshifts \citep{dekel20_flip}. We thus expect that LRDs will disappear once their host halo masses exceed $\Mv \ssim 10^{11}\msun$. This turns out to be the mass of 2-sigma peak halos at $z \ssim 4$.

%==============================
\subsection{Outline}

A parallel study (Dutta Chowdhury et al. in preparation) is underway, which performs N-body simulations of the dry evolution of a galactic disk with FFB clusters and BH seeds, and investigates the formation of the resultant compact central stellar system and the growth of a SMBH.

\smallskip
The current paper is organized as follows. In \se{clusters} we summarize the properties of fiducial cosmic-dawn disks and the FFB clusters in them. In \se{dry_migration} to \se{dry_tide} we evaluate analytically the possible formation of LRD-like compact central stellar systems via dry inward migration and mergers of FFB clusters: in \se{dry_migration} we estimate the migration timescales for two-body segregation and dynamical friction, in \se{dry_mergers} we evaluate the compactness of the merger-built central stellar system, and in \se{dry_tide} we work out the mutual tidal effects of the clusters during the migration process. Then, in \se{BH}, we address the growth of SMBHs via mergers of BH seeds which are carried in by the migrating clusters, and highlight the need for further deepening of the central galactic potential wells in order to overcome SMBH ejection by GW recoils. In \se{compaction} we use a variety of cosmological simulations to study the deepening of the potential wells by wet compaction events due to wet mergers, and their contribution to the formation of compact central stellar systems. In \se{epoch} we address the favored epoch, masses and abundance of LRDs. Finally, in \se{conc}, we summarize our results and conclude.

%%%%%%%%%%%%%%%%%%%%%%%%%%%%%%%%% 2
\section{FFB Clusters}
\label{sec:clusters}

\smallskip % examples FFB
To be used as an example in our estimates below, we recall the properties of a fiducial FFB galaxy at $z \seq 9$ according to \citet{dekel23}. The dark-matter halo virial mass, radius and velocity are
\be
\Mv \simeq 6\stimes 10^{10}\msun\, (1+z)_{10}^{-6.2} \, , 
\ee
\be
\Rv \simeq 12.3\kpc\, \Mveight^{1/3}\, (1+z)_{10}^{-1} \, ,
\ee 
\be
\Vv \simeq 148\kms\, \Mveight^{1/3}\, (1+z)_{10}^{1/2} \, ,
\ee
where $\Mv \sequiv 10^{10.8}\msun\, \Mveight$ and $(1+z) \sequiv 10\,(1+z)_{10}$. With a global star-formation efficiency of $\epsilon \seq 0.3 \epsilon_{0.3}$ \citep{li24}, the total stellar mass in the galactic disk, all initially consisting  of clusters, is
\be
\Md \simeq 3\times 10^9\msun\, \epsilon_{0.3} \, .
\ee
With a contraction factor ($\sim$ halo spin parameter) $\lambda \sequiv 0.025\lambda_{.025}$, the half-mass radius of the disk is
\be
\Rd \simeq 307\pc\, \lambda_{.025}\, . 
\ee

\smallskip % M(Re)
Within $\Rd$, the dark-matter halo mass is roughly $M_{\rm halo}(\Rd)\ssimeq (\Rd/\Rv)\Mv \simeq 1.6\stimes 10^9\msun\lambda_{.025}$. The stellar mass is comparable, $0.5\Md \simeq 1.5\stimes 10^9\msun \epsilon_{0.3}$. The total mass within $\Rd$ is therefore 
\be
M(\Rd) \simeq 3 \times 10^9\msun \, .
\ee
The circular velocity at $\Rd$, due to the halo and added disk, is thus
\be
V(\Rd) \simeq \left( \frac{G M(\Rd)}{\Rd} \right)^{1/2} \simeq 206 \kms \, .
\label{eq:VRe}
\ee

\smallskip % clusters
The FFB disk is predicted to consist of star clusters, in the mass range $10^4 \sdash 10^7\msun$. These limits are determined by the minimum mass shielded against feedback from earlier generations of clusters and by the Toomre mass, respectively \citep[][\S 7]{dekel23}. The cluster mass function can be approximated by $\dd N/\dd m \sprop m^{-\alpha}$ with $\alpha \slsim 2$. This is based on zoom-in cosmological simulations \citep{mandelker14,mandelker17}, and is a generic result in a supersonic turbulent medium \citep{hopkins13,trujillo19,gronke22}. The typical radius of a $10^6\msun$ clump is $R_{\rm c} \ssim 7\pc$. The corresponding densities are on the order of $n \ssim 10^4 \cmc$ and $\Sigma \ssim 3 \stimes 10^3 \msun\pc^{-2}$, above the thresholds for FFB.

%%%%%%%%%%%%%%%%%%%%%%%%%
\section{Dry Migration of Clusters: Timescales}
\label{sec:dry_migration}

A galactic disk that consists of star clusters, embedded in a dark-matter halo, will evolve by gravity into a compact central stellar system. This evolution can be driven by two-body segregation and by dynamical friction. Using analytic toy models, we estimate here the timescales for these processes to bring the clusters from the disk half-mass radius $\Rd$ to the center. The migrating clusters will bring in with them the seed BHs that formed in the cluster centers (\se{BH}). These crude estimates are to be quantified using N-body simulations (Dutta Chowdhury et al., in preparation).

%===============================
\subsection{Two-body segregation}
\label{sec:seg}

% 2-body
For a system of objects with a distribution of masses, two-body relaxation causes segregation, making the massive objects migrate to the center. Assuming that the clusters largely remain intact as point masses (see \se{dry_tide}), and very crudely assuming a spherical system for this toy-model estimate, the segregation timescale at radial distance $r$ can be approximated by \citep[e.g.][]{giersz94}
\be
\tseg \sim \frac{0.17\, N}{\ln(0.11\, N)} \, \td \, .
\label{eq:tseg}
\ee
Here $\td$ is the dynamical time at $r$, 
\be 
\td \simeq (4\pi G \rho)^{-1/2} \simeq \frac{r}{V} \, ,
\label{eq:td}
\ee
with $\rho$ the density at $r$, and $V$ the circular velocity there. The second equality is valid, e.g., for an isothermal sphere. In particular, we will refer to $\td$ at the stellar half-mass radius $\Rd$. For equal-mass objects of mass $\Mc$ in a system of total mass $\Md$ the number of objects relevant for two-body relaxation inside the half-mass radius is $N \seq 0.5\,\Md /\Mc$. For a multi-mass system, $N$ should be replaced by $N \seq 0.5\,\Md/M_{\rm c,max}$, where $M_{\rm c,max}$ is the typical mass of the most massive objects. This has been estimated by \citet{spitzer71} and \citet{portegies04} for core collapse in a star cluster, where the stars are distributed according to a standard \citet{kroupa01} initial mass function, namely of negative slope $\alpha \seq 2.3$ at large masses. This estimate has been confirmed in simulations \citep{rizzuto21,rantala24}. Similar estimates were obtained by \citet{portegies02,devecchi09} as a function of the average object mass. For a population of clusters, the mass function is expected to be somewhat flatter, $\dd N/\dd \Mc \sprop \Mc^{-\alpha}$ with $\alpha \slsim 2.0$ \citep{mandelker14,mandelker17}, but we assume that the estimate of the effective $N$ based on $M_{\rm c,max}$ should be qualitatively valid. We note that the rotation of the disk has been neglected here and this gives a qualitative conservative estimate. In presence of rotation, due to {\it gravo-gyro} instability, the two-body relaxation can be expected to be faster by a factor of $\sim 2-10$ depending on the level of rotation \citep[e.g.,][]{kim08, hong13, kamlah22}.

\smallskip % FFB
For a fiducial FFB galaxy (\se{clusters}), with a disk mass $\Md \seq 3\stimes 10^9\msun$ ($\times \epsilon_{0.3}$) and clusters of maximum mass $M_{\rm c,max} \sequiv 10^6\msun \, M_{\rm c,max,6}$, the effective number of objects is $N \seq 3\stimes 10^3\, M_{\rm c,max,6}^{-1}$. With $V(\Rd) \simeq 200\kms$ and $\Rd \simeq 300\pc$ ($\times \lambda_{.025}$), the dynamical time at $\Rd$ is
\be
\td(\Rd) = \frac{\Rd}{V} \sim 1.5\Myr \, .
\label{eq:td}
\ee
Then, from \equ{tseg}, the segregation time, for $M_{\rm c,max} \seq 10^6 \sdash 10^7\msun$ respectively, is approximately
\be
\tseg \sim (132-22) \Myr \, .
\label{eq:tseg_ffb}
\ee
For given FFB halo properties, $\tseg \sprop \epsilon_{0.3}\,\lambda_{.025}$.

\smallskip
\Fig{dry_mig} shows the segregation timescale as a function of $\Mc$ for the fiducial FFB galaxy and clusters. This estimate of $\tseg$ implies that the two-body segregation could bring the massive clusters from the disk half-mass radius to the galaxy center in a few orbital times, as long as a significant fraction of the clusters remain intact against mutual tidal stripping (see \se{dry_tide}). We note that the cluster-resolving simulation by \citet[][]{chen26} indeed reveals cluster mass segregation and inward migration on a timescale of $\sim\!10\Myr$.

%=============================
\subsection{Dynamical Friction}
\label{sec:df}

% DF
While dynamical friction is understood as a complex non-local process involving resonances \citep{tremaine84,kaur18,banik21}, a useful qualitative approximation can be obtained via the classical more local \citet{chandrasekhar43} formalism, which sums up the two-body interactions between the moving massive object and the much less massive particles in the medium. The smooth component of the stellar disk, combined with the dark-matter halo, exert dynamical friction on the orbiting clusters, each hosting a black hole, which helps bringing them into the galaxy center. We next evaluate the contributions of the halo and the disk to the dynamical-friction-driven migration of the clusters.

%-------------------------
\subsubsection{Dynamical friction by the halo}
\label{sec:df_halo}

% halo DF
Based on \citet{chandrasekhar43}, crudely applied to a spherical isothermal system, the dynamical-friction deceleration is
\be
a_{\rm h} \simeq  F_{\rm h}\, 4\pi\, G^2\, 
\frac{\rho_{\rm h}\,\Mc}{V_{\rm rel}^2} \, .
\label{eq:a_h1}
\ee
Here $\rho_{\rm h}$ is the dark-matter density in the halo at $r$, $\Mc$ is the object mass, and $V_{\rm rel}$ is the relative velocity between the object and the medium which we identify here with the total circular velocity given by $V^2 \seq G M(r)/r$. The factor $F_{\rm h}$ is of order unity, defined by $F_{\rm h} \sequiv {\cal B}\,\ln\Lambda$, where
\be
{\cal B}(x) = {\rm erf}(x)- \frac{2x}{\sqrt{\pi}} e^{-x^2}\, ,
\quad
x = \frac{V_{\rm rel}}{\sqrt{2}\sigma} \, ,
\quad
\Lambda \ssim \frac{M(r)}{\Mc}\, ,
\ee
with $\ln \Lambda$ the Coulomb logarithm. It is derived by integrating over the distribution of orbit impact parameters, between $b_{\rm max} \ssim r$ and $b_{\rm min}\ssim G\Mc/V^2$.

\smallskip  % using td
Using $V^2 = G M(r)/r$, the deceleration becomes 
\be
a_{\rm h} \simeq F_{\rm h}\, 4\pi\,G\,\rho_{\rm h} \, r \frac{\Mc}{M(r)}
= F_{\rm h}\, \frac{V_{\rm h}}{t_{\rm h}(r)}\, \frac{\Mc}{M(r)} \, .
\label{eq:a_h2}
\ee
The second equality is obtained by expressing the halo dynamical time $t_{\rm h}$ at $r$ using \equ{td} for the quantities contributed by the halo alone, $\rho_{\rm h}$ and $V_{\rm h}$.

\smallskip % with 0.5 Md = Mh
Assuming that the halo mass and disk mass within $\Rd$ are comparable, $0.5\,\Md \ssim M_{\rm h}(\Rd)$, as in the fiducial FFB galaxy, we have $V_{\rm h} \ssim V/\sqrt{2}$ and $t_{\rm h} \ssim \sqrt{2}\,\td$, so we finally estimate at $\Rd$
\be
a_{\rm h} \sim \frac{F_{\rm h}}{2} \, \frac{V}{\td}\, \frac{\Mc}{M(\Rd)} \, .
\label{eq:a_h4}
\ee
Recall that $V$, $\td$ and $M$ include the contributions from both the halo and the disk.

\smallskip % tdf
The migration timescale by dynamical friction from radius $r$ to the center can be estimated as \citep[][eq.~8.13]{bt08}
\be
\tdf \simeq \frac{V}{2 a} \seq \frac{V^2}{2\, r\, a}\, \td
\seq \frac{G M(r)}{2\, r^2\, a}\, \td \, ,
\label{eq:tdf}
\ee
where $a$ is the dynamical-friction deceleration at $r$ and $V$ is the circular velocity there.

\begin{figure} % 1
\centering
\includegraphics[width=0.47\textwidth]
{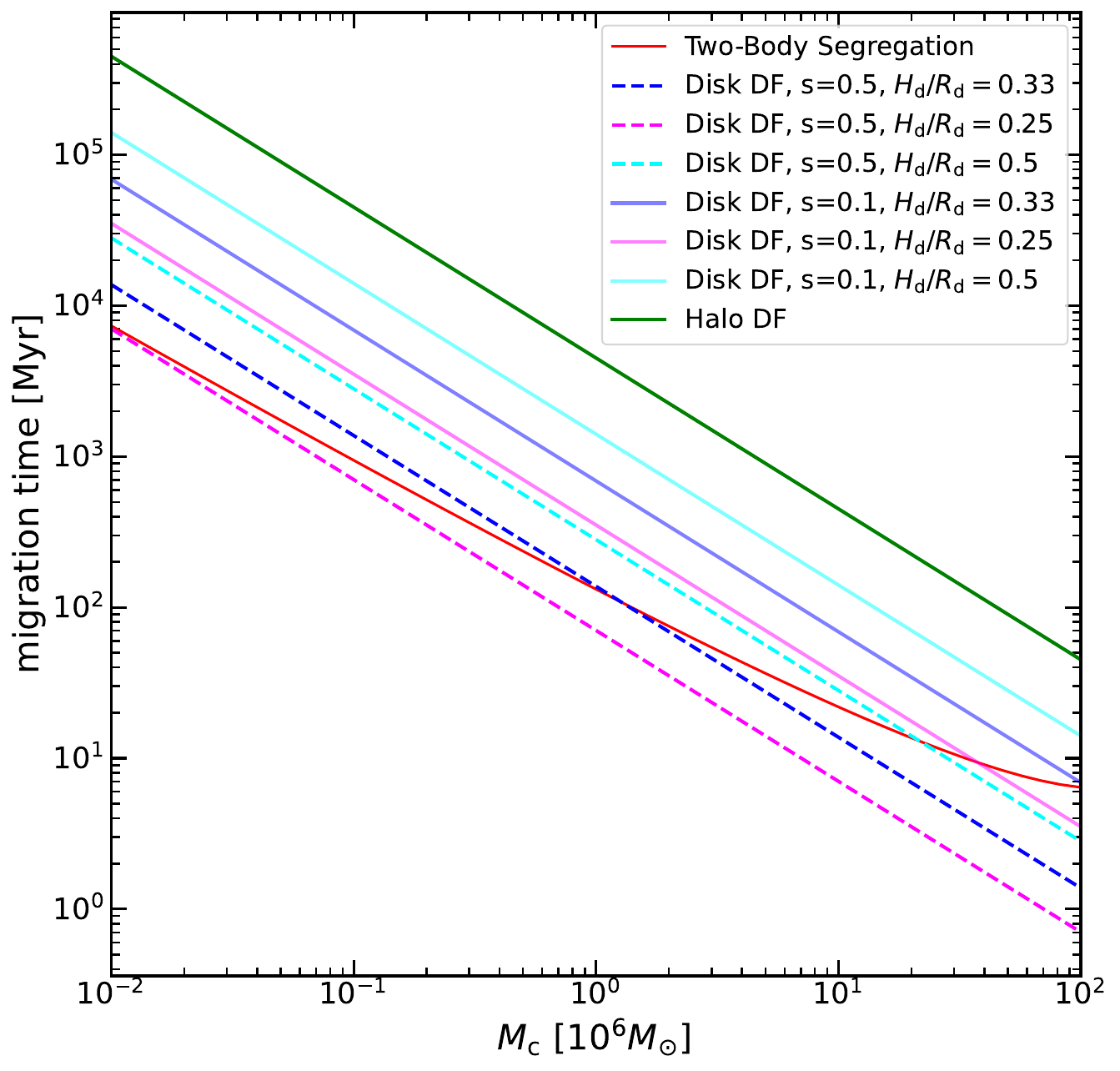}
\caption{Migration timescales from the half-mass radius of the disk to the center as a function of cluster mass. Shown are timescales for two-body segregation (red, \equ{tseg}), for dynamical friction by a hot disk (\equ{tdf_disk_1}) for a smoothed component fraction $s\seq 0.1$ (solid) and $s\seq 0.5$ (dashed), with varying values of the disk aspect ratio, $H_{\rm d}/R_{\rm d}$ = 0.33 (blue), 0.25 (magenta) and 0.5 (cyan), and  for dynamical friction by the halo (green, \equ{tdf_h1}). s=0.5 and  $H_{\rm d}/R_{\rm d}$ = 0.33 are the fiducial values chosen in this study. For segregation, $\Mc$ stands for the maximum cluster mass, while for dynamical friction it is the mass weighted average cluster mass.}
\label{fig:dry_mig}
\end{figure}

\smallskip % FFB
By inserting $a_{\rm h}$ from \equ{a_h4} in \equ{tdf}, we obtain for the timescale associated with the dynamical friction
contributed by the halo at $\Rd$, for a fiducial FFB galaxy where $0.5\,\Md \ssim M_{\rm h}(\Rd)$, 
\be
t_{\rm df,h} \sim F_{\rm h}^{-1}\, \frac{M(\Rd)}{\Mc}\, \td \, . 
\label{eq:tdf_h1}
\ee
%\smallskip % fiducial FFB
For a numerical estimate, assuming a fiducial FFB galaxy, with $M(\Rd) \ssimeq 3\stimes 10^9\msun$ and $\td \simeq 1.5\Myr$,
consisting of clusters of mass $\Mc \seq 10^6\msun\, M_{\rm c,6}$, and assuming $F_{\rm h} \ssim 1$, we obtain 
\be
t_{\rm df,h} \sim 4.5\Gyr\, M_{\rm c,6}^{-1} \, .
\label{eq:tdf_h2}
\ee
This is shown as a function of $\Mc$ in \Fig{dry_mig}. We learn that it is much longer than the timescale associated with two-body segregation, \equ{tseg_ffb}, as well as with the timescale associated with dynamical friction by the disk once it has a significant smooth component (see next). We can thus practically neglect the contribution of dynamical friction by the halo.

%---------------------
\subsubsection{Dynamical friction by the smooth disk}
\label{sec:df_disk}

% disk DF
The dynamical friction exerted on an object rotating in a disk by the rotating background particles is expected to be more efficient than in a spherical system. As evaluated in \citet{dekel25_bh}, this is largely because the relative velocity of the object with respect to the background particles, $V_{\rm rel}$, is comparable to the velocity dispersion $\sigma$ rather than the rotation velocity $V$. Since the dynamical-friction deceleration is proportional to $V_{\rm rel}^{-2}$, the deceleration in a disk, of a density similar to the halo density, is expected to be larger by a factor $(V/\sigma)^2$, with an additional contribution by a geometrical factor. 

\smallskip
\citet[][eqs.~19 and 21]{dekel25_bh} evaluated the dynamical-friction deceleration in a disk to be
\be
a_{\rm d} \simeq F_{\rm d}\, 8\pi\, G\, \Sigma_{\rm s}'(r)\, r\, 
\left( \frac{\Mc}{M(r)} \right)^\mu\, \left( \frac{\Rd}{\Hd} \right)^\nu  \, .
\label{eq:a_d1}
\ee
For a kinematically hot thick disk (or when $\Mc \sll \Md$), which is relevant at high redshifts, the power indices are $\mu \seq 1$ and $\nu \seq 2$. For a cold thin disk (or when $\Mc$ is not much smaller than $\Md$), they are $\mu \seq 2/3$ and $\nu \seq 1$. The ratio $\Hd/\Rd$ is the relative thickness of the disk, which can be approximated by $\Hd/\Rd \ssim \sigma/V$. The dependence of $a_{\rm d}$ on $(\Hd/\Rd)^2$ reflects the mentioned dependence on $\sigma^{-2}$ rather than on $V^{-2}$. The factor $F_{\rm d}$ is of order unity; its value is uncertain and can be calibrated using simulations \citep[e.g.][]{dekel25_bh}.

\smallskip % Sigma
In \equ{a_d1}, $\Sigma_{\rm s}(r)$ is the surface density of the smooth component of the disk, and $\Sigma_{\rm s}'(r)$ is its radial derivative. For an order-of-magnitude estimate, the quantity $\Sigma_{\rm s}'(r)\, r$ can be approximated by $\Sigma_{\rm s}(r)$ in the main body of the disk. This is given that for an exponential disk they are equal at the exponential radius $r_{\rm exp}$, and that the half-mass radius is in the same ball park, $\Rd \ssimeq 1.68\, r_{\rm exp}$. In the main body of the disk, we also crudely replace $\Sigma(r)$ with its average inside $r$, $\bar{\Sigma}(r)$, the latter being higher by $43\%$ at $r_{\rm exp}$ for an exponential disk.

\smallskip % Sigma s
We assume that the smooth component of the disk mass at the time relevant for dynamical friction is a fraction $s$ of the stellar disk mass,
\be
M_{\rm d,s} \seq s\,\Md \, ,
\ee
which we apply within $\Rd$. Then $\bar{\Sigma}_{\rm s} \seq s\,\bar{\Sigma}$ for the disk, so \equ{a_d1} becomes
\be
a_{\rm d} \simeq F_{\rm d}\, 8\pi\, G\, s\, \bar{\Sigma}(r)\,
\left( \frac{\Mc}{M(r)} \right)^\mu\, \left( \frac{\Rd}{\Hd} \right)^\nu  \, .
\label{eq:a_d2}
\ee
At the early stages $s$ is small, as estimated analytically in \se{dry_tide} and by simulations in \citet{dekel25_bh}, so the disk contribution to the dynamical friction is minor. As the migration continues, the clusters tidally strip each other outskirts to form a smooth stellar background that grows inside out from the galaxy central region on a timescale of several tens of Megayears, making the dynamical friction more effective. We adopt $s \seq 0.5$ as the fiducial value, relevant for the masses within $\Rd$ after several tens of Megayears.

\smallskip % tdf
Translating the deceleration to a dynamical-friction timescale using \equ{tdf}, and using $\bar{\Sigma}(\Rd) \seq 0.5\Md/(\pi \Rd^2)$ in \equ{a_d2}, we obtain
\be
t_{\rm df,d} \simeq \frac{1}{8 F_{\rm d}}\, 
s^{-1} \,
\left( \frac{M(\Rd)}{\Md} \right) \,
\left( \frac{M(\Rd)}{\Mc} \right)^\mu \, 
\left( \frac{\Hd}{\Rd} \right)^\nu \, 
\td \, .
\label{eq:tdf_disk_1}
\ee 

\smallskip
For a fiducial FFB galaxy, where $M(\Rd) \ssim \Md$, assuming $\Md \seq 3\stimes 10^9\msun$, $\Mc \seq 10^6\msun\,M_{\rm c,6}$, $\td \seq 1.5\Myr$, $s \seq 0.5$, $\Hd/\Rd \seq 1/3$ in a hot disk ($\mu\seq 1$ and $\nu\seq 2$), and $F_{\rm d} \seq 1$, we finally obtain
\be
t_{\rm df,d} \sim 124\Myr\, M_{\rm c,6}^{-1} \, .
\label{eq:tdf_disk_3_ffb}
\ee 
The disk dynamical friction timescale is shown as a function of $\Mc$ for different values of $s$ and $H_{\rm d}/R_{\rm d}$ in \Fig{dry_mig}.

\smallskip % Compare disk to halo
Comparing the dynamical-friction decelerations by the disk and by the halo, using \equ{a_d2} and \equ{a_h2}, we obtain
\be
\frac{a_{\rm d}}{a_{\rm h}} \simeq 2\, \frac{F_{\rm d}}{F_{\rm h}}\,
\frac{s\,\bar{\Sigma}(\Rd)}{\rho_{\rm h}\,\Rd} 
\left( \frac{\Rd}{Hd} \right)^\nu \, .
\ee
With $\rho_{\rm h}\,r \seq \bar{\Sigma}_{\rm h}(r)/4$ for an isothermal sphere, and $\bar{\Sigma}_{\rm h}(\Rd) \seq \bar{\Sigma}(\Rd)$ for the fiducial FFB galaxy, this becomes
\be
\frac{a_{\rm d}}{a_{\rm h}} \sim 8\, \frac{F_{\rm d}}{F_{\rm h}}\,
s\, \left( \frac{\Rd}{\Hd} \right)^\nu \, .
\ee
With $F_{\rm d} \seq F_{\rm h}$, $s \seq 0.5$, $\Rd/\Hd \seq 3$, and $\nu \seq 2$, the deceleration by the disk is larger than the halo contribution by a factor of $36$.  Even given the uncertainties in the numerical factors $F$, and the crude approximations made in the analysis, it is clear that, once $s$ is a non-negligible fraction of unity,
the halo contribution to the dynamical friction is negligible compared to the disk contribution.

\smallskip % compare to seg
We learn that, once $s$ is a significant fraction of unity, and for reasonable values of the disk aspect ratio $H_{\rm d}/R_{\rm d}$, the timescale for dynamical friction by the disk, as estimated for a fiducial FFB galaxy in \equ{tdf_disk_3_ffb}, is comparable to the two-body segregation timescale, as estimated by \equ{tseg_ffb}. Both are on the order of $100\Myr\, M_{\rm c,6}^{-1}$. Clusters more massive than $10^6\msun$ can migrate inwards in less than $100\Myr$. We therefore expect cluster migration and the emergence of a significant central stellar system on this timescale, as a result of mergers between the migrating clusters.

%===============================
\subsection{Migration summary}

\Fig{dry_mig} summarizes the different timescales for migration from the disk half-mass radius to the center as a function of cluster mass for typical FFB galaxies. They are computed for a fiducial FFB disk of clusters (\se{clusters}). Shown are the timescales for two-body segregation (\equnp{tseg}), for dynamical friction by a hot disk (\equnp{tdf_disk_1}, with $\Fd\seq 1$) with a smoothed component fraction $s\seq 0.1$ and $s\seq 0.5$, and for dynamical friction by the halo (\equnp{tdf_h1}, with $\Fh\seq 1$). Note that in the segregation timescale $\Mc$ stands for the maximum cluster mass, while in the dynamical friction estimates it refers to a certain average cluster mass. We learn that the two-body segregation timescale dominates the migration. For $\Mc \ssim 10^6\msun$ it is estimated to bring half the clusters to the center in $\sim\!100\Myr$. Clusters of higher maximum mass of $\sim\!10^7\msun$ will get to the center in $\sim\!20\Myr$.The dynamical friction by the disk operates on a longer timescale in the early stages of the migration when $s$ is small. If and when the stripped mass into a smooth component reaches half the disk mass, $s \ssim 0.5$ (see \se{dry_tide} for estimates of $s$), the dynamical friction timescale becomes comparable to the segregation timescale. The dynamical friction by the halo is estimated to provide a less important contribution, operating on a timescale of a few Gigayears.

\smallskip % BHs
Seed black holes that form at the centers of the clusters will be carried by the inspiraling clusters and brought to the center where they could be ready to merge. However, we note that according to \citet[][\S 3, Fig.~2]{dekel25_bh}, the very massive clusters of $\sim\!10^7\msun$, unlike the $\sleq 10^6\msun$ clusters, are not expected to form black-hole seeds very efficiently. It is the clusters of $\sim\!10^6\msun$ that are more relevant for the BH growth (see \se{BH}).

\smallskip
The migration timescale of a few tens of Megayears is on the order of the duration expected for the FFB phase at cosmic dawn, and is much shorter than the $\sim\!1\Gyr$ period available for LRD formation at cosmic morning. We thus expect the onset of LRDs by dry migration and mergers to start soon after the end of the FFB phase at $z \ssim 8$.

\begin{figure*} % 2
\centering
\includegraphics[width=0.99\textwidth]
{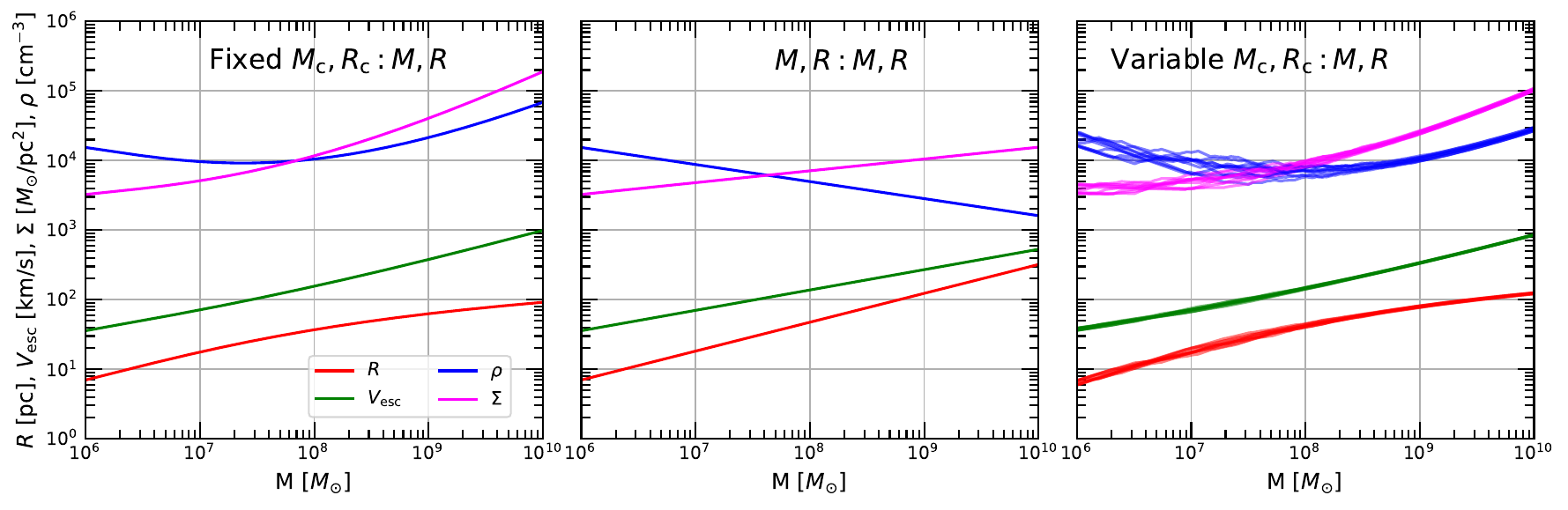}
\caption{Evolution of central dry merger remnant. A sequence of mergers is followed using \equ{merger}. {\bf Left:} the incoming clusters are of mass $\Mc \seq 10^6\msun$. {\bf Middle:} the incoming clusters are of mass $M$ equal to the current mass of the central system. {\bf Right:} the incoming clusters of mass $\Mc$ are drawn from a power law mass function with a slope of $1.5$, maximum mass = $10^7 \msun$, and minimum mass = $10^4 \msun$. The cluster half-mass radii are obtained from the FFB threshold surface density of $3.25 \times 10^3 \msun \pc^{-2}$ within them. Shown are the half mass radius $R$, the corresponding 3D and surface mean densities within $R$, $\rho$ and $\Sigma$ for the central growing merger remnant, and the escape velocity (multiplied by $\sqrt{2}$ to account for a dark-matter halo of similar mass within R).}
\label{fig:dry_mergers}
\end{figure*}

%%%%%%%%%%%%%%%%%%%%%%%%%%%%%%%%%%%%%%
\section{Growth of a Central System by Dry Mergers}
\label{sec:dry_mergers}

We next estimate the compactness of the final central merger remnant, starting from a population of clusters, each with a given mass and radius, that migrate to the galaxy center and merge there. For each dry binary merger, we use the simple recipe by \citet{covington08}, which has been calibrated by a variety of binary merger simulations. They demonstrated that in a dry binary merger of systems of mass $M_1$ and $M_2$ with half-mass radii $R_1$ and $R_2$, respectively, the final radius of the merger remnant $R_{\rm f}$ is well approximated by
\be
\frac{(M_1+M_2)^2}{2\,R_{\rm f}} = \frac{M_1^2}{2\,R_1} + \frac{M_2^2}{2\,R_2} 
+\frac{M_1\,M_2}{R_1+R_2} \, .
\label{eq:merger}
\ee
This equation reflects conservation of energy of virialized systems, with the addition of the mutual potential energy of the two systems when their centers  are at a distance $R_1+R_2$. Similar estimates were obtained by others \citep{cole00,hatton03,shen03}. For an equal-mass merger, with initial half-mass radii $R_{\rm i}$, this gives
\be 
R_{\rm f} = \frac{4}{3} \,R_{\rm i} \, .
\label{eq:Rf_equal}
\ee
For more minor mergers, the growth of radius for a given growth of mass is smaller.

\smallskip % clusters
We ``simulate" the growth of the central merger remnant by a sequence of binary mergers with incoming clusters. Each cluster is assumed to be of mass $\Mc \sequiv 10^6\msun\, \Mcsix$ and half-mass radius $\Rc \sequiv 7 \pc\, \Rcseven$. We note that $R_{\rm c}$ is the birth radius of the cluster and we ignore any evolution in it during the core collapse and formation of the seed BH. The corresponding mean densities within $\Rc$ are
\be
\rho_{\rm c} \seq 1.54 \stimes 10^4\cmc\, \Mcsix\,\Rcseven^{-3} \, ,
\ee
\be
\Sigma_{\rm c} \seq 3.25\stimes 10^3 \msun \pc^{-2}\, \Mcsix\,\Rcseven^{-2} \, .
\ee
Here and later (\se{compaction}, \se{conc}), $\rho$ is expressed in units of molecular hydrogen mass = $2 m_{\rm p}$, where $m_{\rm p}$ is the proton mass. The escape velocity, 
$V_{\rm esc}^2 \seq 2\, V_{\rm c}^2 \seq 2\, G\, (0.5\, \Mc)/\Rc$, is
\be
V_{\rm esc} \seq 25.3\kms\, \Mcsix^{1/2}\,\Rcseven^{-1/2} \, .
\ee

\smallskip % mergers
We denote the growing mass of the central stellar system by $M$ and its half-mass radius by $R$. In a merger, the mass $M$ becomes $M_1+M_2$, and the post-merger radius is given by \equ{merger}. \Fig{dry_mergers} shows the evolution of the central stellar system half-mass radius $R$ and the mean densities within it as a function of the growing $M$ for different merger scenarios. The left panel shows mergers between the central stellar system of mass $M_1 \seq M$ and an incoming cluster of mass $M_2 \seq 10^6 \msun$, radius $R_2 = 7 \pc$, while the middle panel highlights a sequence of mergers between identical systems, where $M_2 \seq M$ and $R_2=R$. Finally, in the right panel, incoming clusters of mass $M_2 \seq M_{\rm c}$ drawn from a power law mass function of slope $1.5$, maximum mass = $10^7 \msun$, and minimum mass = $10^4 \msun$ merge with $M_1 \seq M$ ($10$ different Monte-Carlo realizations are shown). The radii of the incoming clusters are varied accordingly, such that their surface density is fixed at $3.25 \times 10^3 \msun \pc^{-2}$ \citep[the threshold for FFB clusters according to][]{dekel23}. Also shown in \Fig{dry_mergers} is the escape velocity, $\sqrt{2}\,V(R)$, multiplied by an additional factor of $\sqrt{2}$ to account for a dark-matter halo of comparable mass within $R$. The dark matter halo mass is crudely approximated in this way to give a simple order of magnitude estimate rather than calculating the escape velocity from the exact enclosed mass within the growing $R$.

\smallskip % results
The sequence of mergers with mass $\Mc$ and $M$ leads to higher densities than the sequence with equal-mass mergers. For example, starting from $\Mc \seq 10^6\msun$ and $\Rc \seq 7\pc$, when the central mass $M$ reaches $10^9\msun$, its properties are
\be
R = 62.7\pc \, , 
\ee
\be
\rho = 2.14\times 10^4\cmc \, , \quad
\Sigma = 4.05\times 10^4\Msun\pc^{-2} \, , 
\ee
\be
V_{\rm esc} = 378\kms \, .
\ee
The obtained central properties are similar when starting with more massive clusters of $\Mc \seq 10^7\msun$ and $\Rc \seq 15\pc$ (not shown) or with a cluster mass function and variable cluster radii. In the variable $\Mc$, $\Rc$ setup, when the central mass $M$ reaches $10^9\msun$, we get
\be
R = 80\pc \, , 
\ee
\be
\rho = 10^4\cmc \, , \quad
\Sigma = 2.5\times 10^4\Msun\pc^{-2} \, , 
\ee
\be
V_{\rm esc} = 350\kms \, .
\ee
Similar results (not shown) are also obtained for the canonical mass function slope of $\sim 2$.

\smallskip % overestimate stripping
We note that this estimated compactness is likely an overestimate because we ignored the mutual tidal stripping of the clusters, which could be significant after a few tens of Megayears (\se{dry_tide}). On the other hand, even a small fraction of gas in the merging systems can enhance the compactness of the remnant, as found in \citet{covington08}, and will be addressed in \se{compaction}.

\smallskip % simulations Dhruba
Dutta Chowdhury et al. (in preparation) performs N-body simulations of the dry evolution of a galactic disk with FFB clusters and BH seeds. In these simulations, after $\sim\!100\Myr$, the average stellar surface density is $\Sigma \ssim 10^6\msun\pc^{-2}$ within $10\pc$ and $\Sigma \sgsim 10^4\msun\pc^{-2}$ within $100\pc$. The corresponding escape velocities at these radii are high, $800\kms$ and $600\kms$, respectively. These are comparable to our analytic estimates in \Fig{dry_mergers}. We caution, however, that these simulations assume a rather top-heavy mass function for the clusters, with a flat slope of $\alpha \seq 1.4$ (instead of a more realistic $\sim\! 2$ slope) and maximum cluster mass as high as $10^8\msun$ (compared to the FFB Jeans mass of $\sim\!10^6\msun$ and Toomre mass of $\sim\!10^7\msun$). This likely overestimates the inward migration efficiency and the resultant compactness of the central cluster. Furthermore, as mentioned, according to Fig.~4 of \citet{dekel25_bh}, the clusters more massive than $\sim\!10^6\msun$ would find it difficult to form seed BHs in $3\Myr$, before the death of the massive stars and the onset of feedback, such that the migration of the massive clusters may not be associated with the growth of a SMBH (see \se{BH}). On the other hand, the estimated stellar masses of a few of the observed high-redshift clusters are possibly as high as $10^8\msun$ \citep{claeyssens23,messa24}, making the high end of the cluster mass function in the simulations acceptable in some cases. Also, certain simulations \citep{vergara25a,vergara25b} indicate more efficient core collapse than in other simulations, possibly allowing seed BH growth in more massive clusters. The caveat is that these simulations assume unrealistic extremely compact initial clusters (see \se{BH}).

\smallskip % LRD. BHs.
The estimated central radius and densities are close to the LRD stellar properties as crudely estimated for a hybrid LRD model of a stellar system with AGN.  For a system of mass $M \sequiv 10^9\msun\, M_9$ inside radius $R \sequiv 100\pc\, R_{100}$, the LRD mean densities within $R$ are
\be
\rho \simeq 4.24 \times 10^4\cmc\, M_9 R_{100}^{-3} 
\left( \frac{R}{H} \right)_3 \, , 
\ee
\be
\Sigma \simeq 3.18 \times 10^5\msun \pc^{-2}\, M_9 R_{100}^{-2} \, ,
\ee
and the escape velocity is
\be
\Vesc \simeq 299\kms\, M_9^{1/2} R_{100}^{-1/2} \, .
\ee
This escape velocity is actually comparable to the value of $291\kms$ estimated for a fiducial FFB galaxy (\equ{VRe}). If this relatively moderate compactness is valid for LRDs, the dry evolution of clusters may be sufficient for reproducing the LRD properties. However, it is possibly not as high as the value needed for retaining the SMBH after gravitational-wave recoils. According to \Fig{bh_recoil} below and \citet[][Fig.~13]{dekel25_bh}, for a hot disk and a slope $\beta \seq 1.5$ for the BH mass function as they arrive at the center, the required escape velocity is $450\sdash 500\kms$.Wet compaction events may be needed in order to increase the escape velocity to the required values (see \se{BH} and \se{compaction}). 

\begin{figure*} % 3
\centering
\includegraphics[width=0.49\textwidth ,trim={1.0cm 5.5cm 1.5cm 4.0cm},clip]
{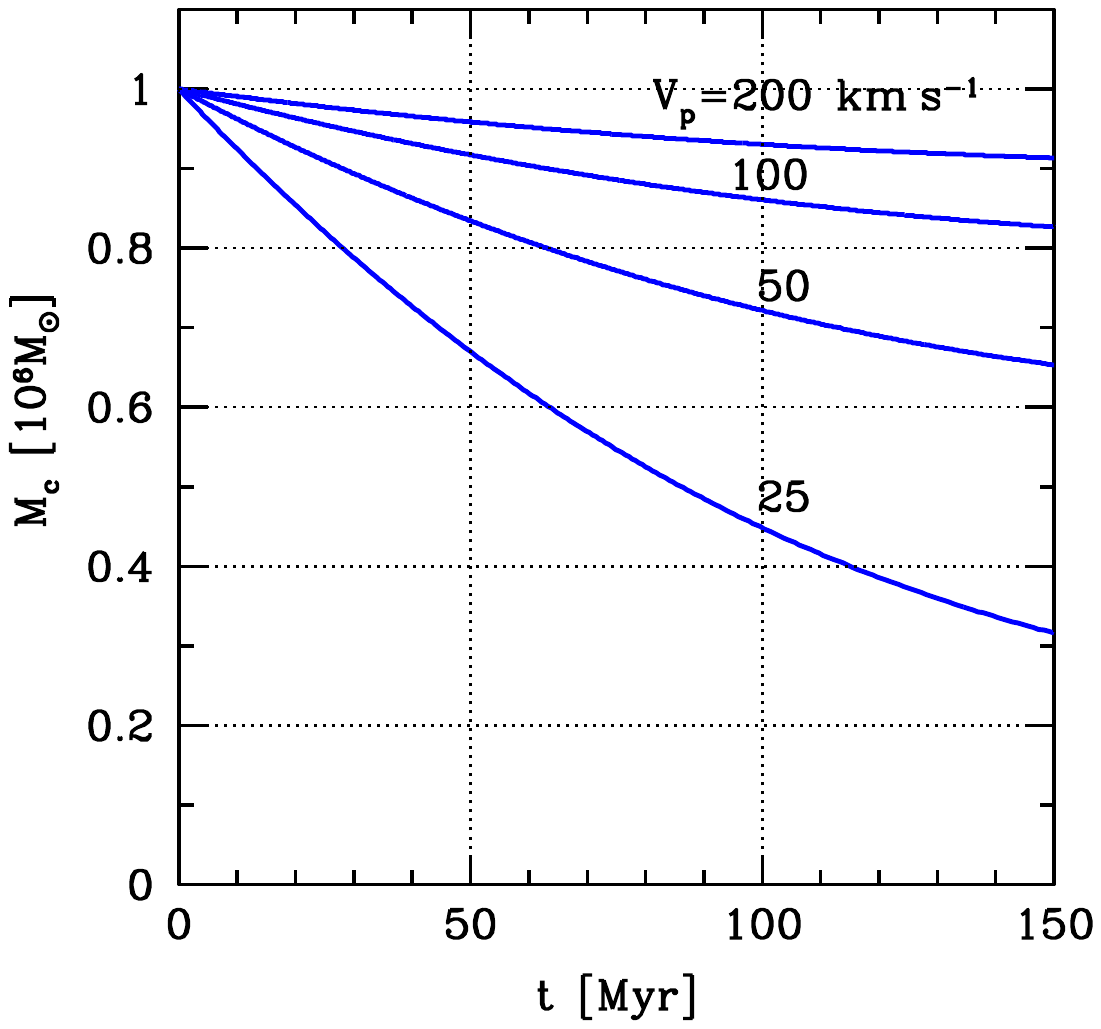}
\includegraphics[width=0.49\textwidth ,trim={1.0cm 5.5cm 1.5cm 4.0cm},clip]
{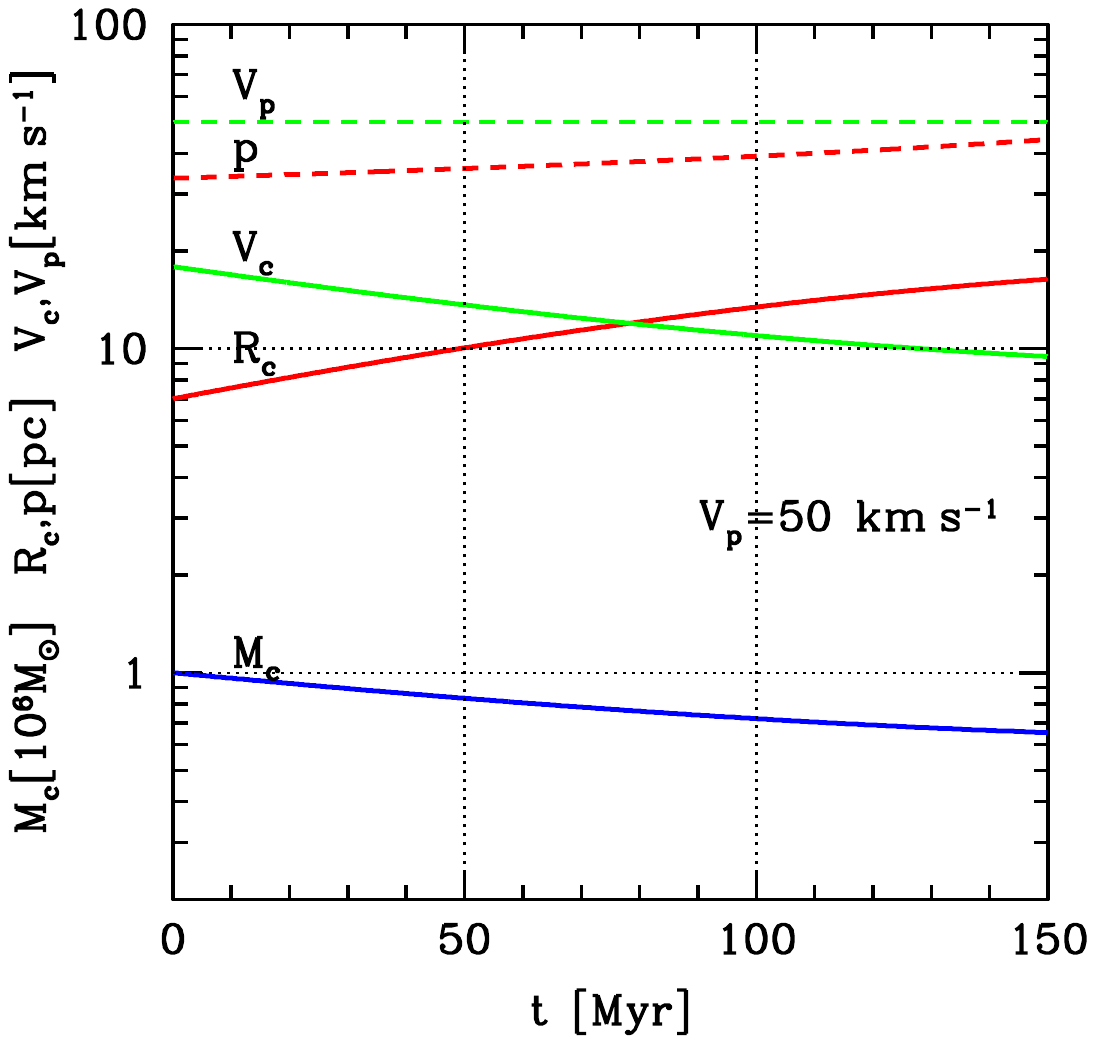}
\caption{Evolution of clusters in a sequence of mutual tidal encounters assuming the impulse approximation and the tidal limit. {\bf Left:} Tidal mass loss, for clusters of initial mass $\Mc \seq \Mp \seq 10^6\msun$ and radius $\Rc \seq 7\pc$, for different values of relative velocities $\Vp$. The clusters largely remain in tact during the migration time of $\sim\!100\Myr$, while a non-negligible smooth component develops for $\Vp \slt 100\kms$. {\bf Right:} Evolution of cluster mass $\Mc$ (blue), radius $\Rc$ (red), and circular velocity $\Vc$ (green), compared to impact parameter $p$ (dashed red) and relative velocity $\Vp$ (dashed green, assumed here to be $50\kms$). The validity of the impulse approximation and the tidal limit are indicated by $\Vp \sgt \Vc$ and $p \sgt \Rc$, respectively.}
\label{fig:dry_tide}
\end{figure*}

%%%%%%%%%%%%%%%%%%%%%%%%%%%%%%
\section{Tidal Effects by Cluster Encounters}
\label{sec:dry_tide}

\def\nut{\nu_{\rm tidal}}

As the clusters orbit and spiral-in within the galactic disk, they gravitationally interact with each other, gain energy and lose mass by mutual tidal encounters. The binding energy and the mass that remains intact in the clusters during their inward migration are key for their migration efficiency, both by two-body segregation (\se{seg}) and by dynamical friction (\se{df}). They are also important for the properties of the merger-driven central system, and for the ability of the clusters to keep the BHs in their centers during the migration. The mass that is stripped away is relevant for the dynamical friction exerted by the disk (\se{df_disk}). In order to address these issues, we crudely estimate the tidal evolution of cluster properties as a function of time.

%==========================
\subsection{The effects of a single encounter}

We follow the analysis of hyperbolic tidal encounters by \citet{dekel80}, based on \citet{spitzer58}. As long as the perturbing clusters orbit with relative velocities $\Vp \ssim 50\kms$ or larger, while the internal velocities inside the clusters are only $\Vc \ssim 10\kms$, one can adopt the {\it impulse approximation}, assuming that the stars in the victim cluster do not move appreciably as the perturbing cluster passes by. We consider a cluster of mass $\Mc$ and rms radius $\Rc$\footnote{For this toy-model estimate we do not distinguish between the rms radius and the half-mass radius.} such that the internal rms velocity is $\Vc^2 = G\Mc/\Rc$. The perturbing cluster is of mass $\Mp$, moving at a velocity $\Vp$ relative to the victim cluster at closest approach that corresponds to an impact parameter $p$. In the {\it tidal limit}, where $p$ is larger than $\Rc$, the rms velocity increment in the cluster can be approximated by 
\citep{spitzer58} 
\be
(\Delta \Vc)_{\rm tidal} \simeq 
\left(\frac{2}{3}\right)^{1/2} \frac{2 G \Mp}{p^2 \Vp} \, \Rc \, .
\ee
The encounter can then be characterized by the dimensionless tidal parameter \citep[][eq.~8]{dekel80} 
\be
\nut \equiv \frac {(\Delta\Vc)_{\rm tidal}}{\Vc} \simeq 
0.10\, \left(\frac{4\,\Vc}{\Vp}\right)\,
\left(\frac{2\Rc}{p}\right)^2\, \left(\frac{\Mp}{\Mc}\right) \, . 
\label{eq:nu}
\ee
We note based on experience that the impulse approximation and the tidal limit are the kind of assumptions that lead to useful approximations even when they are only marginally valid, namely when $\Vp$ and $p$ are not extremely large compared to $\Vc$ and $\Rc$.

\smallskip
For a system with close-to-isotropic stellar orbits, \citet{dekel80} show that the relative energy gain in an encounter is dominated by the secular, second-order term in $\Delta \Vc/\Vc$.\footnote{The first order term, proportional to $\la \Vc \cdot \Delta \Vc \ra$, is expected to vanish for an isotropic system due to the symmetry between ingoing and outgoing stars, but it can become important if radial orbits dominate.} The energy gain of the bound system is calibrated by simulations to be 
\citep[][eq.~10]{dekel80}
\be
\frac{\Delta \Ec}{\Ec} \simeq -\frac{2}{3} \nut^2 \, .
\label{eq:DE}
\ee
The associated relative mass loss is found there to be
\be
\frac{\Delta \Mc}{\Mc} \simeq -\frac{2}{9} \nut^2 \, .
\label{eq:DM}
\ee
Note that $\Delta \Ec$ is positive and $\Delta\Mc$ is negative.

\smallskip % Rc
Assuming that the cluster is in virial equilibrium both before and after the encounter, with corresponding radii $\Rc$ and $\Rc +\Delta \Rc$, masses $\Mc$ and $\Mc + \Delta \Mc$, and circular velocities $\Vc$ and $\Vc + \Delta \Vc$, one can write
\be
1+\frac{\Delta \Rc}{\Rc} = \frac{1 + \Delta \Mc/\Mc}{1 + \Delta \Ec/\Ec} \, ,
\ee
\be
\frac{\Delta \Vc^2}{\Vc^2} = \frac{\Delta \Ec}{\Ec} \, .
\ee
Using \equ{DE} and \equ{DM}, and assuming $\nut \sll 1$, we obtain
\be
\frac{\Delta \Rc}{\Rc}  \simeq \frac{4}{9}\, \nut^2 \, ,
\label{eq:DR}
\ee
\be
\frac{\Delta \Vc^2}{\Vc^2} \simeq  -\frac{2}{3} \nut^2 \, .
\label{eq:DV}
\ee
Thus, the energy, mass, radius and circular velocity of a cluster change as a result of the encounter by a multiplicative factor [$1-(2/3)\nut^2]$, $[1-(2/9)\nut^2]$, $[1+(4/9)\nut^2]$, and $[1-(2/3)\nut^2]^{1/2}$, respectively.

\smallskip % caveat: inter-penetrating
A caveat of the above analysis may refer to the assumption of tidal limit, in cases where $p$ is not sufficiently large compared to $\Rc$. We learn from \citet{dekel80} that for inter-penetrating, more head-on collisions, the more direct velocity increment is $(\Delta \Vc)_{\rm direct} \prop G\Mp/(p\Vp)$, such that $\nut \sprop (p\Vp)^{-1}$ and the actual mass loss is at a lower rate, only $1/20$ of the bound energy change. In this case our estimate of mass loss based on the tidal limit could be an over-estimate of the actual mass loss. Nevertheless, we find in \Fig{dry_tide} below that the tidal limit is expected to be valid in the fiducial circumstances studied.

\smallskip % caveat: 1st order
As mentioned in an earlier footnote, another caveat that should be mentioned for completeness is that for clusters that happened to have more radial stellar orbits, the mass loss would be dominated by the first-order term in $\Delta \Vc/\Vc$ rather then by the second-order term. This is estimated in \citet{dekel80} to be $\Delta\Mc/\Mc \simeq -0.5\,\nut$, such that the mass loss per encounter could be larger by an order of magnitude. We note, however, that the situation of radial orbits is unlikely, given that the FFB clusters form by fragmentation of a rotating disk, and are thus dominated by circular stellar orbits \citep{ceverino12}.

%====================================
\subsection{Evolution under a sequence of encounters}

We next ``simulate" a sequence of binary encounters based on the effects of a single encounter as estimated above. In order to relate the sequence of encounters to time, we estimate the time interval between successive encounters as
\be
\Delta t = (n_{\rm p}\, \sigma\, \Vp)^{-1} \, .
\label{eq:Dt}
\ee
Here the number density of clusters is 
\be
n_{\rm p} = \frac{\Np}{\pi\,\Rd^2\, 2\Hd} \, ,
\label{eq:np}
\ee
where the initial total number of clusters within the half mass radius is 
\be
\Np = \frac{0.5\,\Md}{\Mp} \, , 
\ee
and $\Md$, $\Rd$ and $\Hd$ are the mass, half-mass radius and half thickness of the disk of clusters, respectively. The cross section for an encounter is assumed to be 
\be 
\sigma = \pi\,p^2\, .
\label{eq:sigma}
\ee
% p
The impact parameter is crudely assumed to be the mean separation of neighboring clusters, namely 
\be
p = \Np^{-1/3}\, \left( \frac{\pi\,2\Hd}{\Rd} \right)^{1/3}\, \Rd \, .
\label{eq:p}
\ee

\smallskip % Vp
We consider the relative velocity $\Vp$ to be an input parameter whose value is very uncertain. It depends on the circular velocity, which may increase by contraction and slow down during the inward migration, developing a strong dependence on distance from the center. The value of $\Vp$ will affect the changes per encounter as $\nut^2 \sprop \Vp^{-2}$,
but will counter-affect the encounter rate as $\prop\! \Vp$. The fiducial FFB circular velocity is $\sim\!200\kms$, but the relative velocities $\Vp$ in the rotating disk are likely to be on the order of the velocity dispersion, which is smaller than the circular velocity by a factor of $\Hd/\Rd \ssim 1/3$, namely $\Vp \ssim 67\kms$. The velocity $\Vp$ may become even smaller after the inspiraling clusters slow down by dynamical friction and merging as they approach the center. On the other hand, the galaxy circular velocity may become higher than $200\kms$, either by a factor of order two due to the dry formation of a central compact merger remnant (\se{dry_mergers}) or possibly by a larger factor due to wet compaction events (\se{compaction}). This increase in circular velocity may compensate for the slow down of $\Vp$, leaving fiducial values of $\Vp \ssim 50 \sdash 100\kms$ but with a large uncertainty.

\smallskip % Np
The relevant number of clusters $\Np$ will decrease with time as the clusters gradually migrate and merge into the central cluster. We crudely assume that the disk of clusters is practically evacuated on a timescale that is twice the timescale associated with the inward migration by two-body segregation from $\Rd$ to the center. For a fiducial FFB galaxy and clusters of $10^6\msun$ we estimated in \equ{tseg_ffb} $\tseg \ssimeq 132\Myr$. We very crudely adopt for $t \slt 2\tseg$ a linear decline of
\be
\Np = \Np(t=0)\, \left( 1- \frac{t}{2\tseg} \right) \, .
\label{eq:Npt}
\ee

\begin{figure*} % 4
\centering
\includegraphics[width=0.80\textwidth]{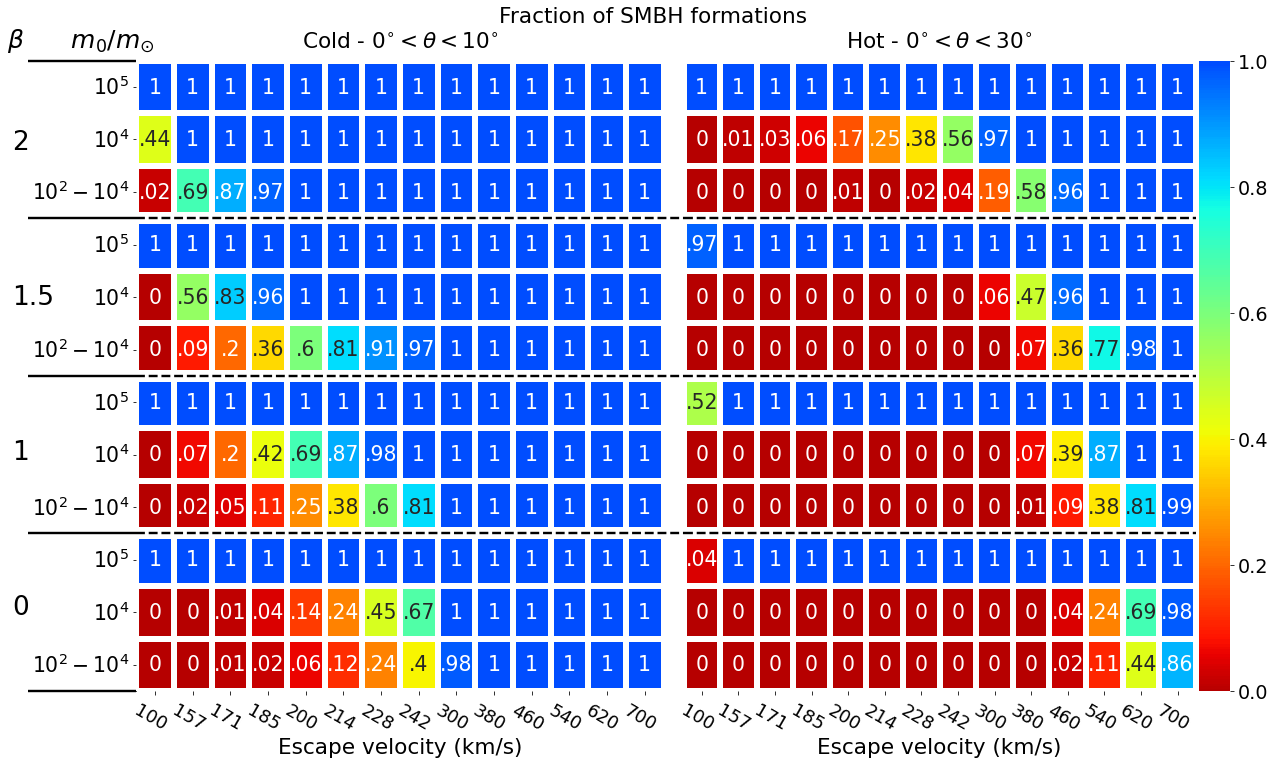}
\caption{The recoil bottleneck of SMBH growth by BH mergers
based on Monte Carlo simulations \citep{dekel25_bh}. The numbers (and colors) in each entry of the table represent the fraction of galaxies that allow SMBH growth exceeding $10^6\msun$, as a function of the escape velocity from the galaxy $\Vesc$. We consider the threshold escape velocity for retaining the SMBH to be given by a fraction of $\sim\!0.5$, namely near the entries colored yellow in the transition from red to blue colors. Shown are four cases of the BH mass-function slope $\beta$, with the fiducial case being $\beta \seq 1.5$. Also shown are cases of cold disk and hot disk, with spin-orbit misalignments $\theta \seq 0 \sdash 10^\circ$ and $\theta \seq 0 \sdash 30^\circ$, respectively. Within each of the eight sub-tables, shown are three cases of the initial primary BH mass, $m_0$, which was either selected to be $10^5\msun$ or $10^4\msun$, or was drawn at random from the seed BH mass function in the range $(10^2 \sdash 10^4\msun)$. Focusing on the fiducial values adopted by \citet{dekel25_bh}, of $\beta\seq 1.5$, a hot disk, and initial BH masses of $(10^2 \sdash 10^4\msun)$, we read that the threshold escape velocity is $\Vesc \ssimeq 487\kms$. In comparison, the fiducial escape velocity of an FFB galaxy is $\Vesc \ssim 200\kms$ before dry mergers of the FFB clusters into a compact central stellar system and wet compaction (\se{clusters}), but it could rise to several hundred $\kms$ after dry mergers (\se{dry_mergers}) and even more after wet compaction (\se{compaction}).}
\label{fig:bh_recoil}
\end{figure*}

\smallskip % changes in each encounter
Before each encounter, we re-compute $\Np$ by \equ{Npt}.The impact parameter $p$ is increased in proportion to $\Np^{-1/3}$ according to \equ{p}.
Then $\np$ and $\sigma$ in \equ{Dt} are changed accordingly following \equ{np} and \equ{sigma}. We equate $\Mp$ with the current $\Mc$, and keep $\Vp$ constant.

\smallskip % Fig results
\Fig{dry_tide} shows the resultant time evolution of the cluster properties in a sequence of encounters where each encounter causes changes based on \equ{DM}, \equ{DR} and \equ{DV}. Focusing on the mass loss, and recalling the uncertainty associated with the value of $\Vp$, the left panel refers to different values of $\Vp$ from $200\kms$ to $25\kms$. The right panel also shows the increase of $\Rc$ and decrease of $\Vc$ in comparison with $p$ and $\Vp$, for $\Vp \seq 50\kms$. The fact that $\Vp$ is significantly larger than $\Vc$ indicates that the impulse approximation is a useful approximation. The fact that $p$ is significantly larger than $\Rc$ indicates that the tidal limit is a useful assumption. We recall that if there were a significant inter-penetration, the effects deduced from the tidal limit would have over-estimated the actual effects.

\smallskip % results intact
We learn from \Fig{dry_tide} that for $\Vp \sgeq 100\kms$ the tidal effects are weak. Even for a relative velocity as low as $\Vp \ssim 25\kms$,
a significant fraction of the mass is still bound to the clusters after the typical migration time of $\sim\!100\Myr$. The survival of bound clusters indicates that our estimated migration timescales (\se{dry_migration}) and merger-driven central cluster properties (\se{dry_mergers}) can serve as reliable toy-model crude estimates.

\smallskip % smooth
The complementary non-negligible fraction of the mass at $\sim\!100\Myr$ is in a stripped, smooth, stellar disk capable of exerting dynamical friction as assumed in \se{df_disk}. This will accelerate the inward migration rate that is driven by two-body segregation, especially at its late stages.

\smallskip % sims bh
The estimated mass loss in \Fig{dry_tide} is in the ball park of the findings in an N-body simulation \citep[][Figs.~5 and 6]{dekel25_bh}.

%%%%%%%%%%%%%%%%%%%%%%%%%%%%%%%%% 
\section{Black Hole Growth}
\label{sec:BH} 

% BH seeds: $\Mc \ssim 10^6\msun$ for core collapse.
As described in \citet{dekel25_bh}, based on several other studies, the FFB clusters of $\sim\!10^6\msun$ at cosmic dawn are natural sites for the formation of BH seeds of $\sim 10^4\msun$ via rapid core collapse to a central very massive star (VMS). The core collapse is sped up to $\slt\!3\Myr$ by two mechanisms. First, {\it gravo-thermal} instability \citep{lynden68}, that is enhanced  by two-body segregation in the presence of young massive stars \citep{spitzer71,hachisu79,portegies02,portegies04,devecchi09,katz15,rizzuto21,rantala24}. Second, {\it gravo-gyro} instability \citep{hachisu79,hachisu82,ernst07,kim08,hong13,kamlah22}, due to rotational support of the clusters in the galactic disk configuration \citep{ceverino12}. It has been estimated \citep[][Fig.~4]{dekel25_bh} that BH seeds are expected to form efficiently in FFB clusters of $\sim\!10^6\msun$ and less. Formation of BH seeds in $\sim\!10^7\msun$ clusters, the estimated Toomre mass in FFB galactic disks, is likely to be efficient only if the stellar IMF is very top-heavy. This is because, otherwise, the core collapse in these clusters is expected to take more than $3\Myr$, allowing the massive stars to die and stop contributing to the speed up of core collapse. Supernova feedback that turns on after $3\Myr$ may suppress the core collapse even further. Following \citet[][]{dekel25_bh}, we thus assume hereafter a maximum cluster mass of $\sim\!10^6\msun$ (and a minimum of $\sim\!10^4\msun$) and a seed black hole of mass $0.01 M_{\rm c}$ at the center of each cluster.

\smallskip % vergara
We note that N-body simulations of clusters \citep{vergara25a,vergara25b} indicate core collapse to a VMS and possibly to a massive seed BH which is more efficient than in the other simulations mentioned above. These results, if extrapolated to clusters more massive than the $\sim\!10^5\msun$ simulated, may apparently allow BH seed growth also in massive clusters. However, the initial clusters assumed in these simulations are extremely compact, with unrealistically small half-mass radii of $(0.005 \sdash 0.05) \pc$.

\smallskip % Cluster-assisted migration.
As estimated in \se{dry_migration}, as long as the clusters are largely intact \citep[see simulations in][Fig.~5]{dekel25_bh}, the cluster migration inwards in the galactic disk can bring a significant fraction of the stellar mass (about $\sim 10^9 M_{\odot}$ in the fiducial case) into a compact central stellar system in $\sim\! 100\Myr$. The clusters should carry the BH seeds with them, bound by the cluster potential wells, bringing about $10^7 \msun$ of seed BH mass in the central merger zone for potential BH-BH mergers. Based on \Fig{dry_tide}, $\Vc$ is expected to be reduced by tidal encounters during the migration by only a factor of order two until they merge with the centrally growing stellar system, likely keeping the clusters sufficiently bound for carrying the BHs with them. The cluster migration thus assists the BH migration into the center where they can merge into a SMBH, making the migration more efficient than assumed for naked BHs in \citet{dekel25_bh}. 

\smallskip % Required escape velocity for recoil.  
For the BH seeds to merge into massive BHs that are retained in the galaxy centers, the merger remnants have to overcome the bottle neck introduced by GW recoils \citep{pretorius05,campanelli06,baker06}. The ejection velocities in the case of BHs with comparable masses and non-negligible misaligned spins, as expected in galactic disks that are not unrealistically cold, are typically of several hundred $\kms$. Based on the Monte Carlo simulations of sequences of BH mergers described in Section 5.5 (Figs. 10-13) of \citet{dekel25_bh}, \Fig{bh_recoil} summarizes the required escape velocity for retaining the BH merger remnant in the galaxy. It is an expansion of the results summarized in Fig.~13 of \citet{dekel25_bh}. In each sequence, the potential central massive BH starts from a primary of a given mass and subsequent binary mergers with a secondary BH are assumed to occur with the primary, whose mass and spin are updated after every merger, at the galaxy center with a given escape velocity. The mass of the secondary is drawn at random from a power law mass function. The simulation continues till the total accreted BH mass amounts to $10^7\msun$. In the event that the recoil velocity of the merged remnant becomes larger than the escape velocity, the remnant is ejected, and the sequence restarts from a new primary mass. We refer the reader to Section 5.5 of \citet{dekel25_bh} for more details and worked out examples of the Monte Carlo simulations. The entries (and colors) in \Fig{bh_recoil} represent the fraction of galaxies (out of 500 sequences) that allow SMBH growth exceeding $10^6\msun$, as a function of the escape velocity from the galaxy $\Vesc$ and three other input parameters. We consider the threshold escape velocity for retaining the SMBH to be given by a fraction of $\sim\!0.5$ of the sequences, namely the entries colored yellow in the transition from red to blue colors. Shown are four cases of the slope $\beta$ of the seed BH mass function $\dd N/\dd m \sprop m^{-\beta}$, as the BHs enter the central merger zone. Also shown are cases of cold disk and hot disk, with spin-orbit misalignments drawn at random in the range $\theta \seq 0 \sdash 10^\circ$ and $\theta \seq 0 \sdash 30^\circ$, respectively. Within each of the eight sub-tables, shown are three cases of the initial primary BH mass $m_0$, which was either selected to be $10^5\msun$ or $10^4\msun$, or was drawn at random from the seed BH mass function in the range $(10^2 \sdash 10^4\msun)$. 

\smallskip % BH results
For no BH spins, or fully aligned spin and orbit, there is 100\% growth in all cases considered (not shown in the figure). We read from \Fig{bh_recoil} that for a cold disk ($\theta \seq 0 \sdash 10^\circ$), there is a significant fraction of growing SMBHs for $\Vesc$ values in the ball park of $200\kms$, as in fiducial FFB galaxies. However, for a more realistic hot disk ($\theta \seq 0 \sdash 30^\circ$), the threshold $\Vesc$ is in the range $240 \sdash 640\kms$, well above $200\kms$. In particular, with the fiducial $\beta \seq 1.5$, a relatively hot disk, and a central primary BH of $10^2 \sdash 10^4\msun$, the required escape velocity for retaining the SMBH against recoil is $V_{\rm esc} \sgt 487\kms$. It could become even larger for a thicker disk with larger misalignments. This is significantly larger than the $\Vesc \sim\! 200\kms$ expected in the typical cosmic-dawn galaxies produced in the FFB phase. The required escape velocity may also be larger than the value of $\slt 378\kms$ estimated in \se{dry_mergers} from the buildup of a compact central stellar system by dry mergers between the clusters. It seems that further deepening of the central potential well is required in order to retain the BHs that are ejected by GW recoils. This could be a natural result of wet compaction events (see below).

\smallskip %caveat
We note that if clusters more massive than $\sim\!10^6\msun$ up to $\sim 10^7 \msun$ were present within the same disk mass budget of $3 \times 10^9 \Msun$ with a canonical power law mass function slope close to $2$, then the mass fraction in clusters above $\sim\!10^6\msun$ would be about $33 \%$ (but only $1 \%$ by number), which may not be able to form seed black holes. These clusters would, therefore, not contribute to the central BH growth but can still participate in the build up of the compact central stellar system, deepening its potential. The somewhat reduced total available BH mass would still be sufficient to power the central BH growth to at least $10^6 \Msun$ and would not significantly change any of our results from the Monte Carlo simulations. 

\smallskip
We further note that in reality, it is unlikely that all BH-BH mergers happen at the galaxy center, as assumed in the Monte Carlo simulations. The assembling compact central stellar system has a finite size of $\lesssim 100 \pc$ (see \se{dry_mergers}) and both cluster-cluster and BH-BH mergers may occur within this extended merger zone. If the circular velocity curve within $100 \pc$ is rising and the magnitude of the inward acceleration is decreasing or constant with $r$, then as long as the recoil velocity is less than the escape velocity at $100 \pc$, recoil from within the merger zone will only be able to push the BH merger remnant out to $100 \pc$, after which it will become bound again, fall back on a dynamical friction timescale depending on its current mass, and can re-grow \citep[see Section~5.3 of][]{dekel25_bh}. On the other hand, if the circular velocity curve within $100 \pc$ is declining with $r$, then for recoil velocities less than the escape velocity at $100 \pc$, the BH merger remnant will remain bound (escape velocities within $100 \pc$ are much higher) and continue to grow with some potential wandering. With this in mind, in the next section, we focus on escape velocity enhancement at $\sim 100 \pc$ due to wet compaction events and compare that to the threshold of $\sim 487 {\rm km/s}$ required to overcome the GW recoil bottleneck.

\smallskip % lock NH
A deepened potential well due to wet compaction also leads to better locking and reduced wandering of the bound, growing BH around the galaxy center. Using the NewHorizon simulations \citep{dubois21}, it has been shown by \citet{lapiner21}, e.g., in their Fig.~4 (illustrated in Figs.~3, C1 and C2), that pre-compaction the central BH is wandering at $\sim\!1\kpc$ about the galaxy center until a compaction event (at $z \ssim 2\sdash 3$) locks it to the center. In these simulations, the post-compaction escape velocity at $1\kpc$ ($V_{\rm esc}\seq \sqrt{2}V_{\rm circ}$) ranges from $200\kms$ (galaxy H51), to $500\kms$ (H1, H6) to $1070\kms$ (H18), to $1350\kms$ (H3, H10) and to an extreme value of $3,400\kms$ (H2). The escape velocities that seem to be sufficient for bringing the BH to the center are typically half the above values. It does seem that in most cases a wet compaction is required for locking the BH to the center of the galaxy.

\smallskip % other goodies of wet compaction
Wet compaction events could help the BH growth in two additional ways. First, the deepening of the potential well, associated with a steepening of the density profile, would be useful in preventing suppression of the dynamical-friction-driven inspiral into the galactic center by core stalling \citep{read06,goerdt06,kaur18,banik21}. Second, the gas compaction could assist in overcoming the final parsec problem \citep{begelman80} by providing central gas. A caveat is that wet compaction events bring in gas that could accrete onto the central BH and produce X-rays that are not seen in typical LRDs. Efficient gas depletion into stars and outflows should be considered in
parallel with BH growth in order to explore this issue.

%%%%%%%%%%%%%%%%%%%%%%%%%%%%%%%%% 4
\section{Wet Compaction}
\label{sec:compaction}

Wet compaction events have been studied using a variety of cosmological simulations, using different codes and physical recipes, focusing on periods from cosmic dawn to cosmic noon, without and with black holes. For example, heavily used and currently available are the VELA and FirstLight suites of simulations utilizing the ART code \citep{ceverino14,ceverino17}, which explore cosmic morning with $\sim\!25\pc$ and $\sim\!13\pc$ resolution for the gas, respectively. We have just ran the MAGE simulations (Yao et al., in preparation), which focus on cosmic dawn with $10\pc$ resolution using the RAMSES code \citep{teyssier02}. The NewHorizon simulation, which includes black holes, explores compactions at cosmic noon with a resolution $\sim\!40\pc$ using the RAMSES code. Unfortunately, these simulations do not yet resolve the FFB physical processes at cosmic dawn and the resultant clusters, which are key for the dry growth of the compact central clusters and for the formation of BH seeds. These would require a resolution of order one parsec.

\smallskip
We analyze here sample galaxies from the VELA and MAGE simulations in order to evaluate the relative growth of compact central stellar systems by wet compaction events, and briefly refer to results from the NewHorizon simulations with BHs, but we cannot use these simulations to study the full combined effects of dry growth by cluster migration and global wet compaction events. This combination is likely to lead to higher compactness and escape velocities than are obtained by each process alone, but exploring the combined effects will have to await simulations of higher resolution, which should eventually also incorporate BHs. Preliminary results from the simulation with $\leq\!3\pc$ resolution at cosmic dawn \citep[][]{chen26} indicate that it begins to capture the combined effects of FFB clusters and compaction events, but we defer its analysis to future work.

\begin{figure*} % 5
\centering
\includegraphics[width=0.99\textwidth]{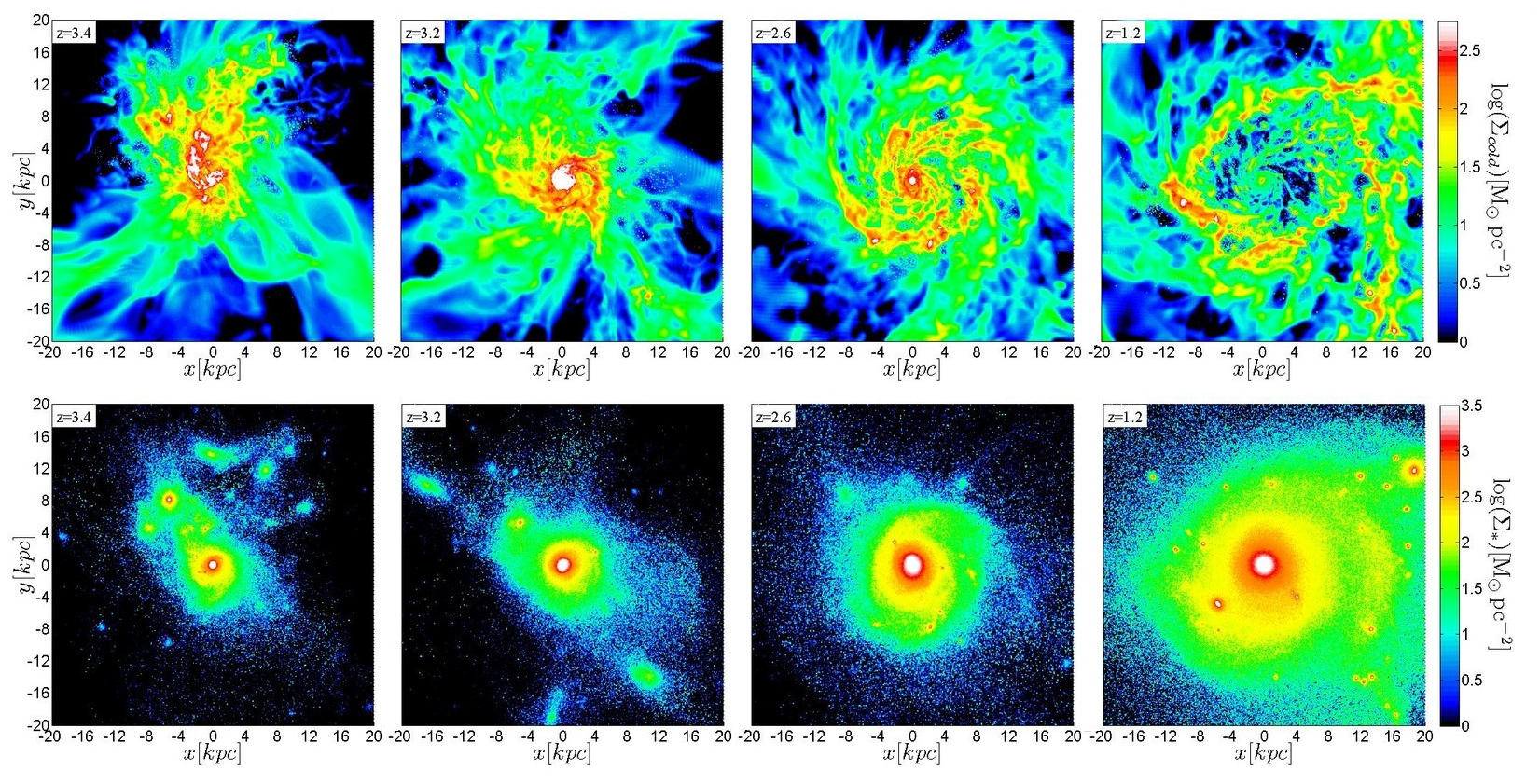}
\caption{Wet compaction in a VELA simulation (V07). Shown is face-on density for gas (top) and stars (bottom). A merger causes gas compaction ($z\seq 3.4$), leading to a compact central star-bursting ``blue nugget" ($z\seq 3.2$), which passively evolves to a long-lived ``red nugget" (from $z\seq 2.6$ to $z\seq 1.2$ and on). The central mass allows the stabilization of an extended star-forming gas disk by incoming streams ($z\seq 2.6$), which evolves into an extended ring ($z \seq 1.2$). The red nugget resembles an LRD, while the emergence of an extended disk marks the end of the phase that is observationally identified as an LRD.}
\label{fig:vela_compaction}
\end{figure*}

\begin{figure*} % 6
\centering
\includegraphics[width=0.60\textwidth]{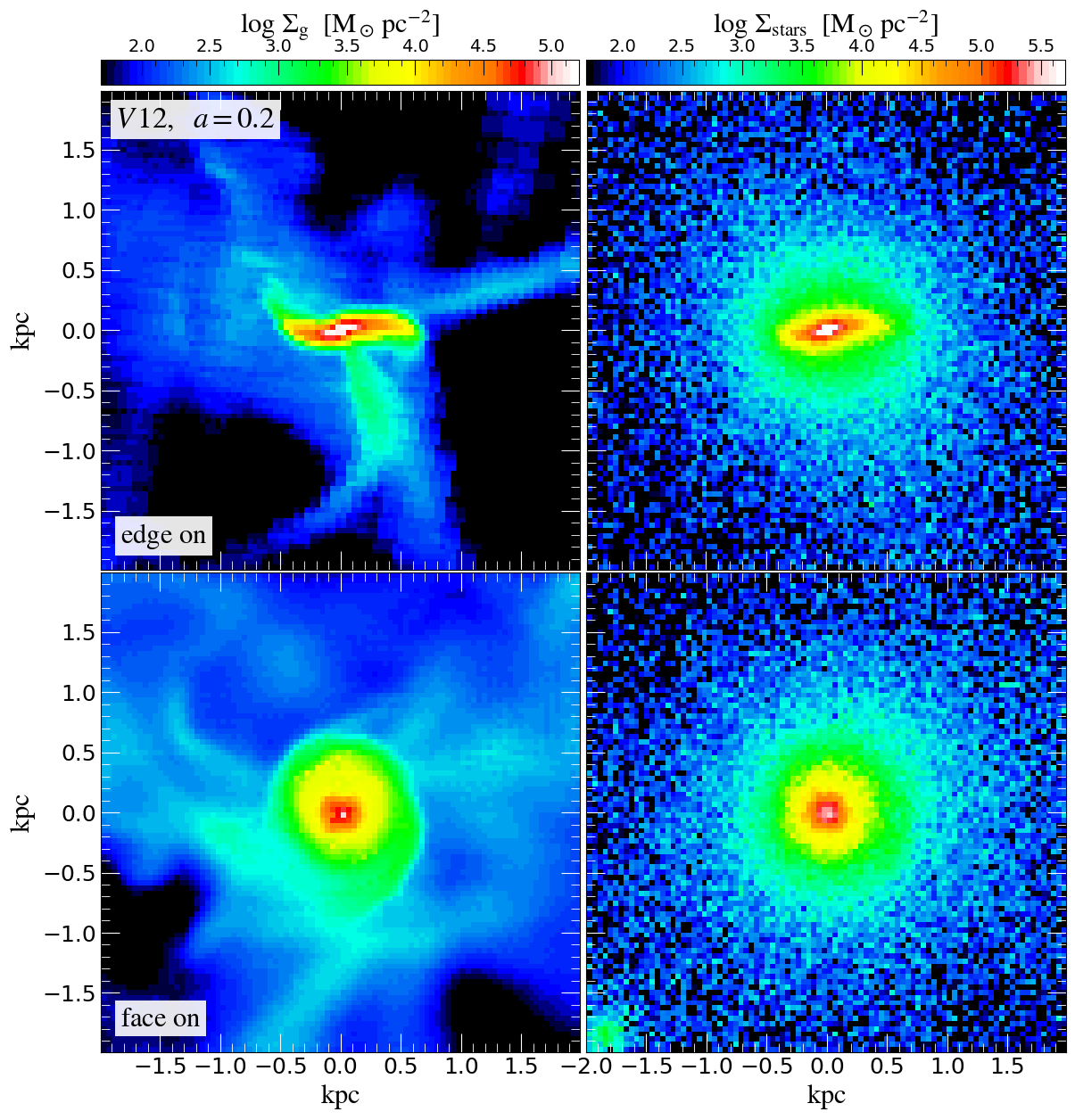}
\caption{Nuclear disk as a result of wet compaction in a VELA simulation (V12) at $z \seq 4$. The mean stellar surface density within a radius of $100\pc$ is in the ball park of $\sim\! 5\stimes 10^5\msun\pc^{-2}$. This implies a mass of $\sim \! 3\stimes 10^{9}\msun$, and an escape velocity $\sim\! 520\kms$.}
\label{fig:vela_nuclear_disk}
\end{figure*}

\begin{figure*} % 7
\centering
\includegraphics[width=0.33\textwidth]
{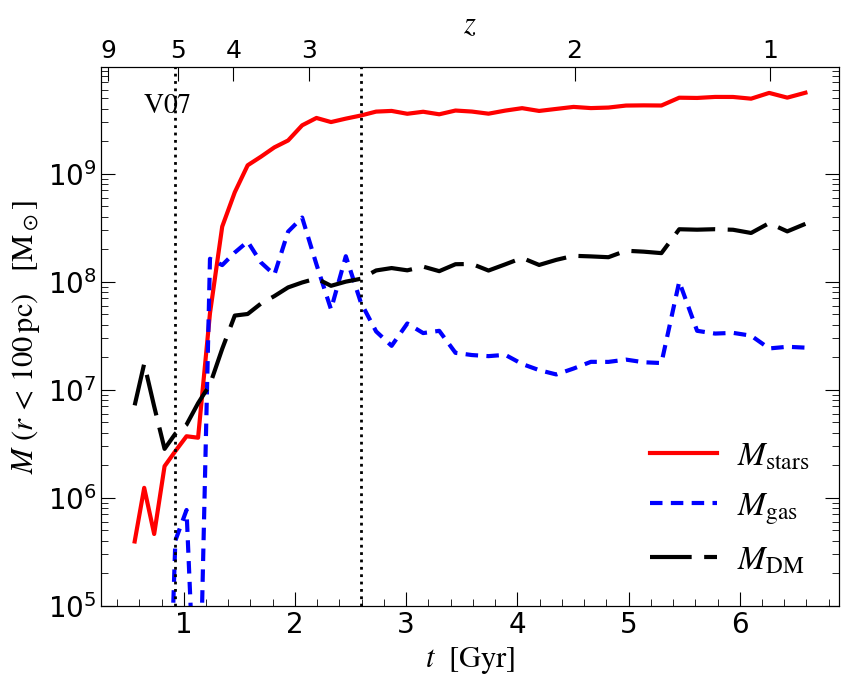}
\includegraphics[width=0.33\textwidth]
{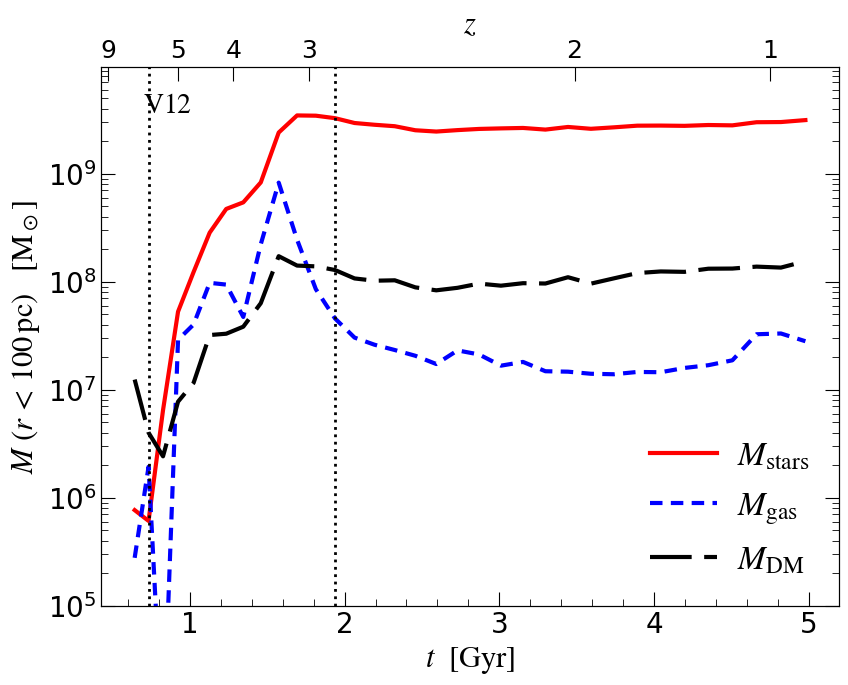}
\includegraphics[width=0.33\textwidth]
{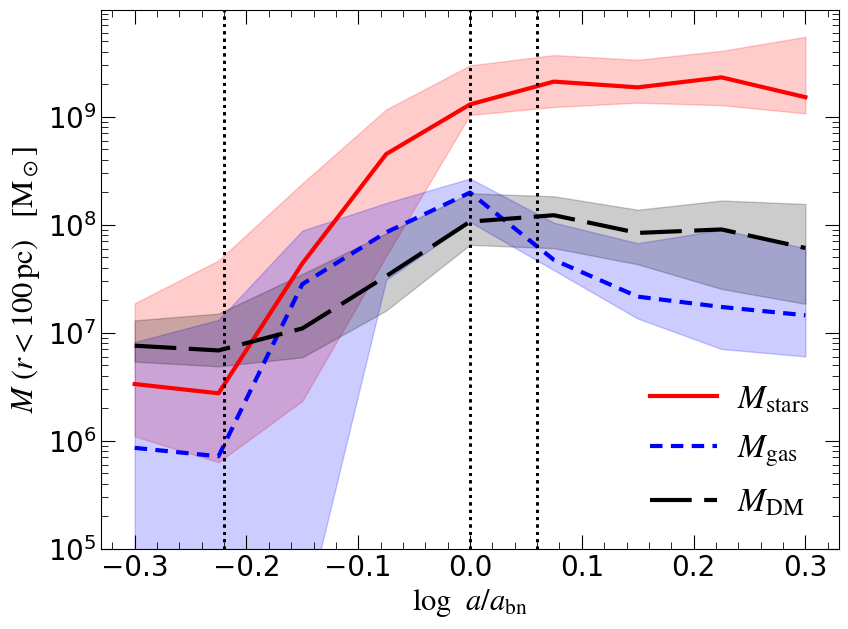}
\includegraphics[width=0.33\textwidth]
{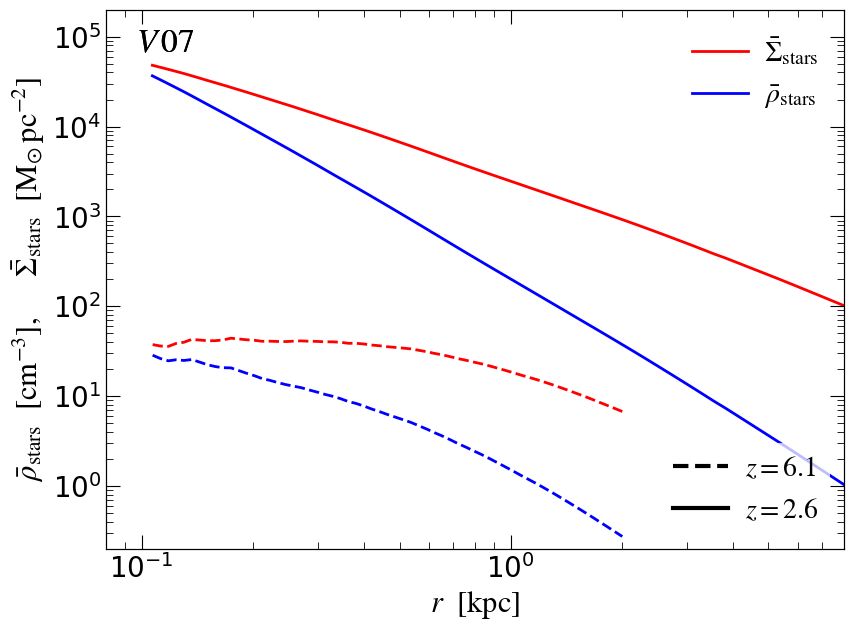}
\includegraphics[width=0.33\textwidth]
{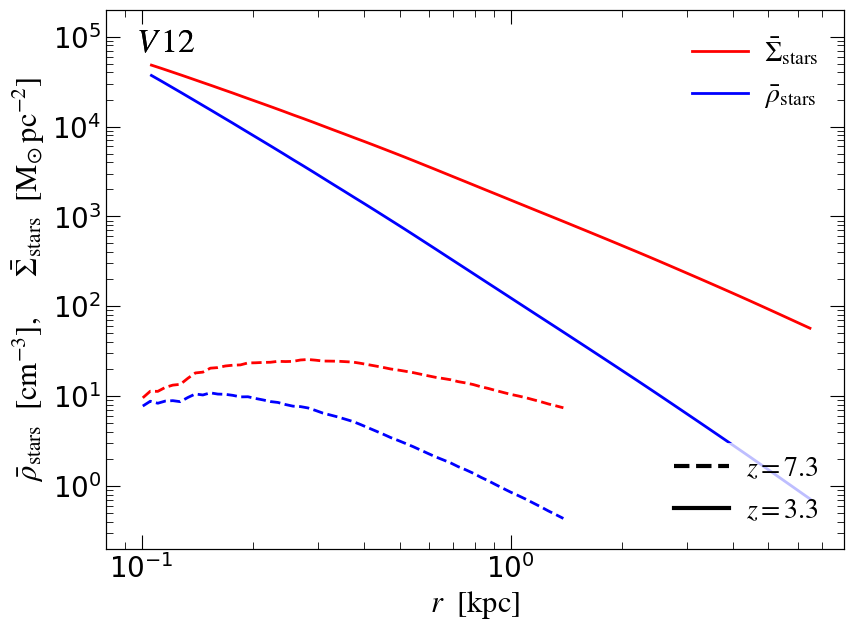}
\includegraphics[width=0.33\textwidth]
{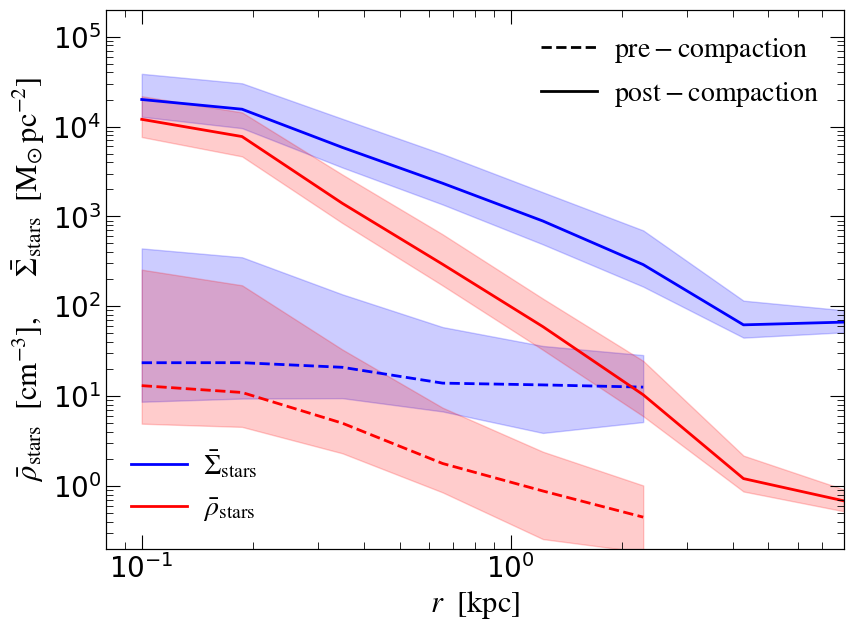}
\includegraphics[width=0.33\textwidth]
{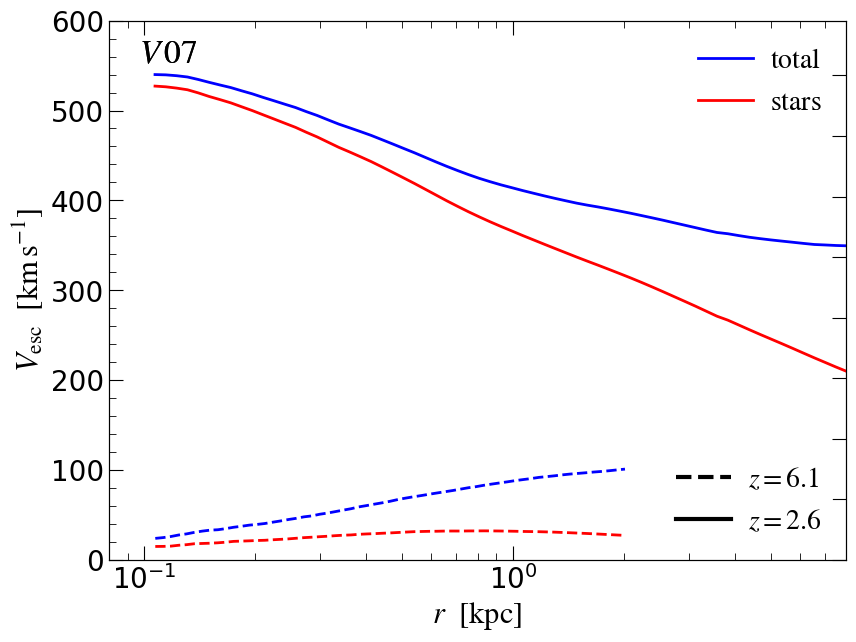}
\includegraphics[width=0.33\textwidth]
{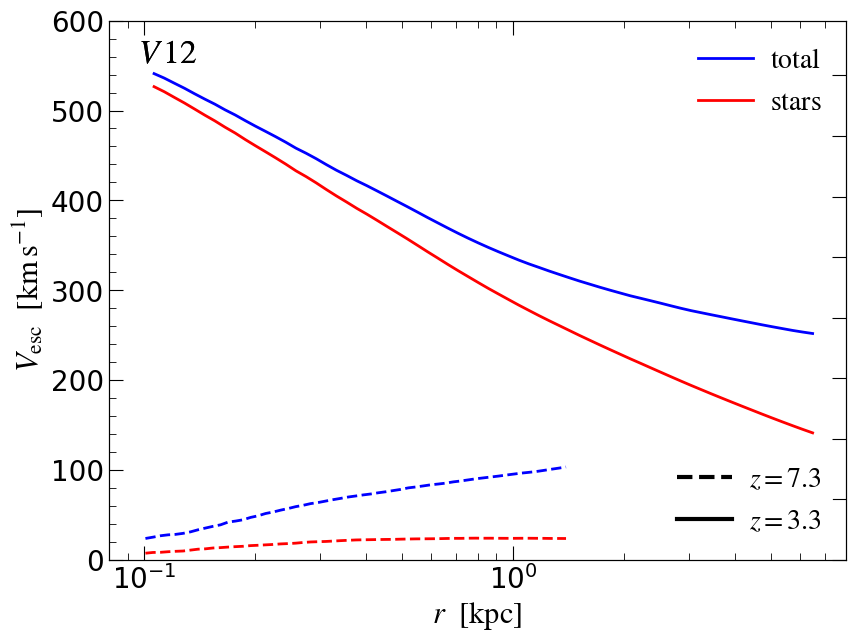}
\includegraphics[width=0.33\textwidth]
{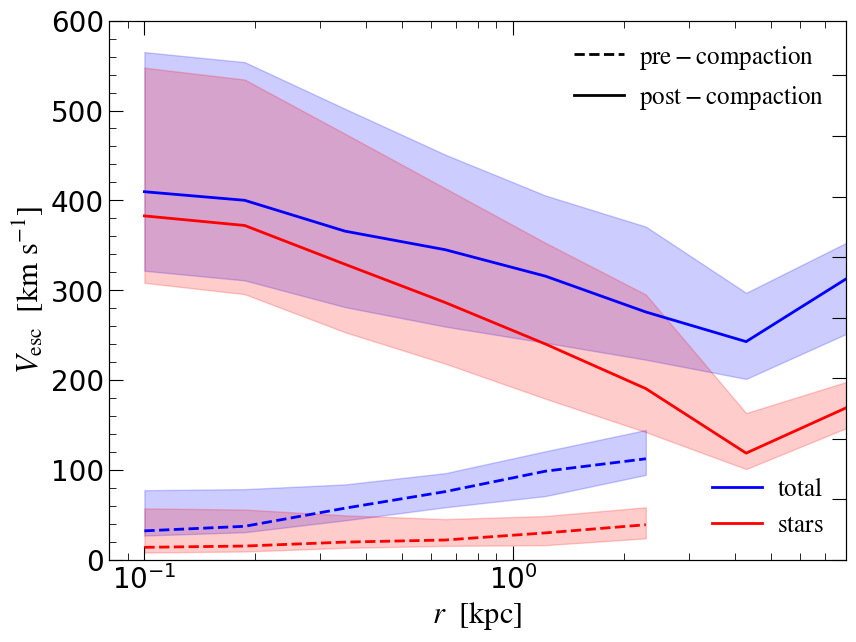}

\caption{Growth of compact central stellar systems by wet compaction events in VELA cosmological simulations at cosmic morning \citep{ceverino14,lapiner23}.
Top: the evolution of mass within $100\pc$ for the stars (red), gas (short-dash blue) and dark matter (long-dash black). Middle: the average stellar density profiles at $<r$ before and after the compaction (dashed and solid). Bottom: the escape velocity at $r$ referring to the total mass inside $r$. The left and middle panels each refer to a major merger event in galaxies V07 and V12, in which the most dramatic compactions are detected. The right panels refer to stacking the 16 most massive galaxies in the VELA suite each shifted horizontally such that the expansion factor $a(t)$ coincides at the time $a_{\rm bn}$, corresponding to the peak of the compaction in terms of gas mass within $100\pc$. Shown are the medians and $\pm 1\sigma$ scatter (shaded). For the stacked profiles, the pre-compaction and post-compaction times are taken to be at $\log(a/a_{\rm bn})=-0.22$ and $+0.06$. The growth of stellar mass within $100\pc$ during the compaction, and the corresponding densities, is typically by $3$dex. The growth of the escape velocity is by $1$dex to $1.5$dex. The post-compaction stellar mass within $100\pc$ reaches $\sim\! 3\stimes 10^9\msun$ with densities $\bar{\rho} \seq (1\sdash 4) \stimes 10^4\cmc$ and $\bar{\Sigma} \seq (2\sdash 5)\stimes 10^4\msun\pc^{-2}$. The escape velocities at $100\pc$ become $(320 \sdash 560)\kms$.} 
\label{fig:vela_prof}
\end{figure*}

\begin{figure*} % 8
\centering 
\includegraphics[width=0.33\textwidth] {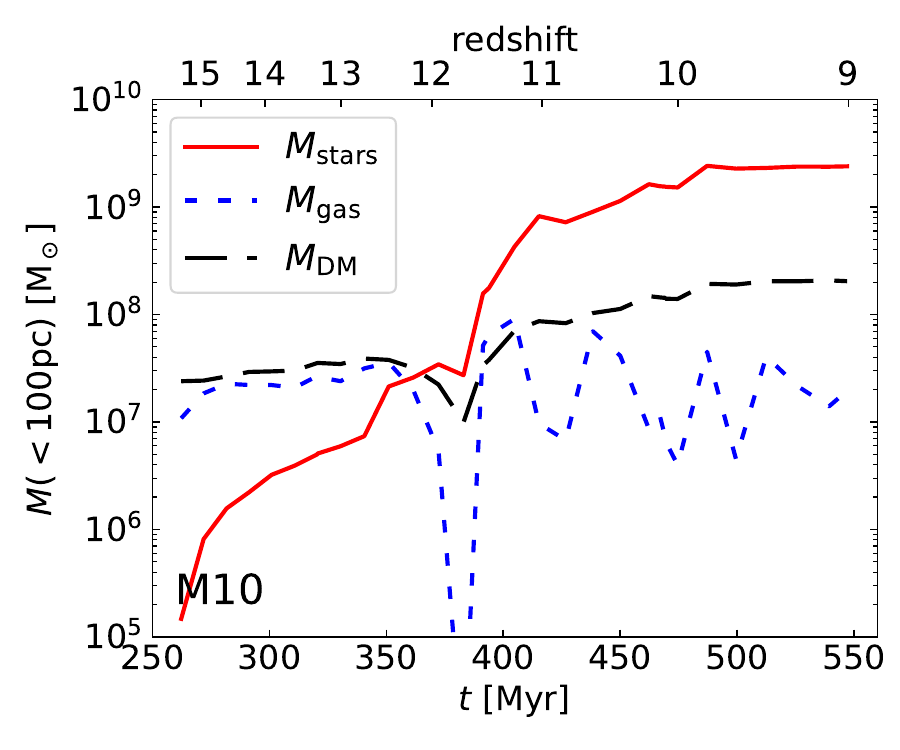}
\includegraphics[width=0.33\textwidth] {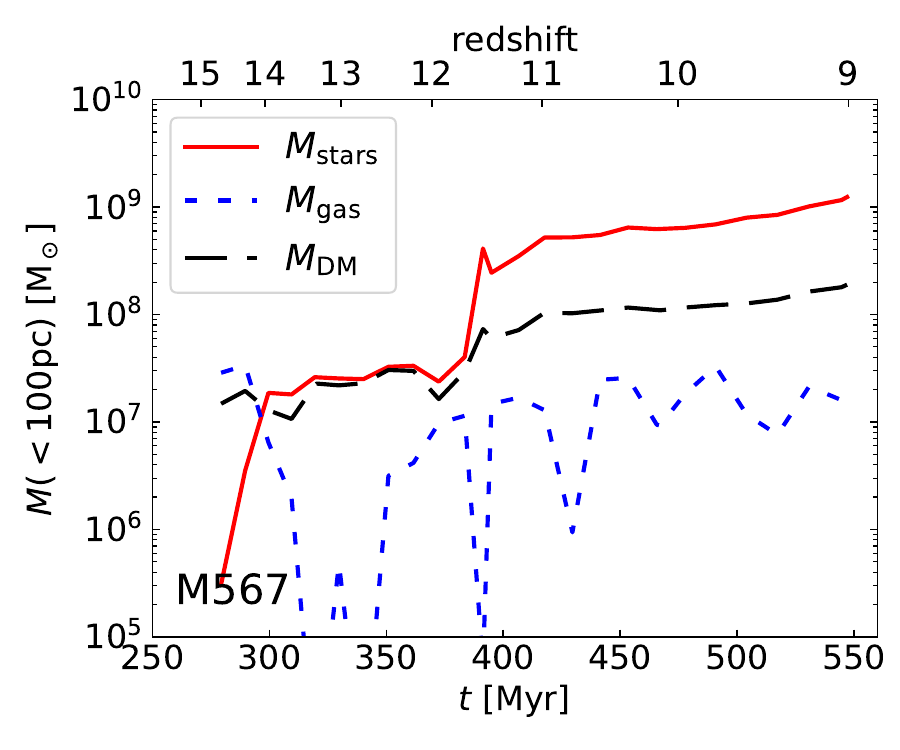}
\includegraphics[width=0.33\textwidth] {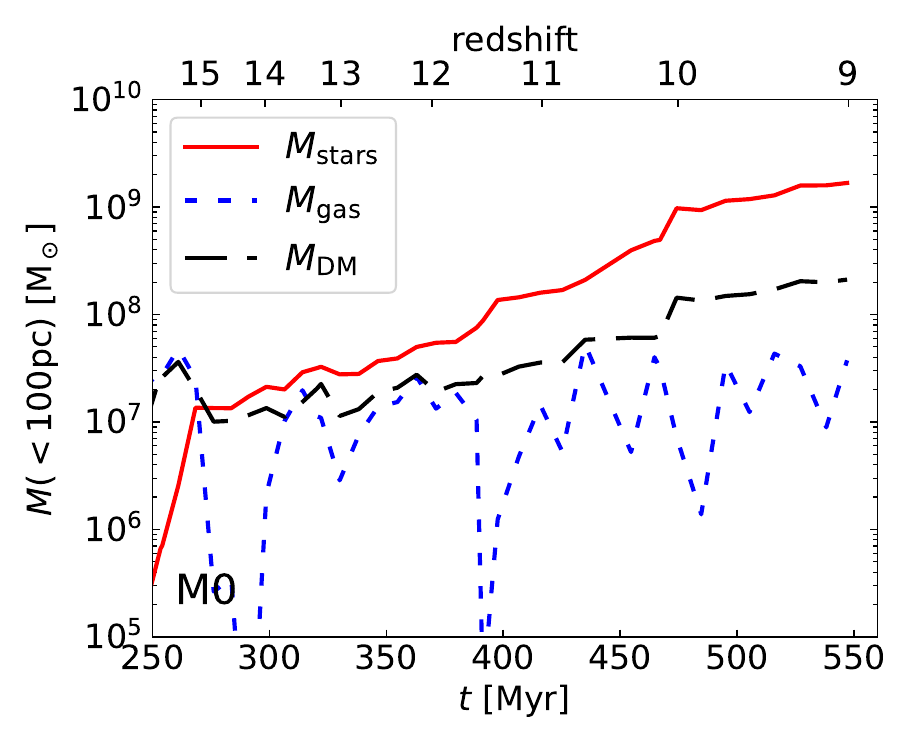}
\includegraphics[width=0.33\textwidth] {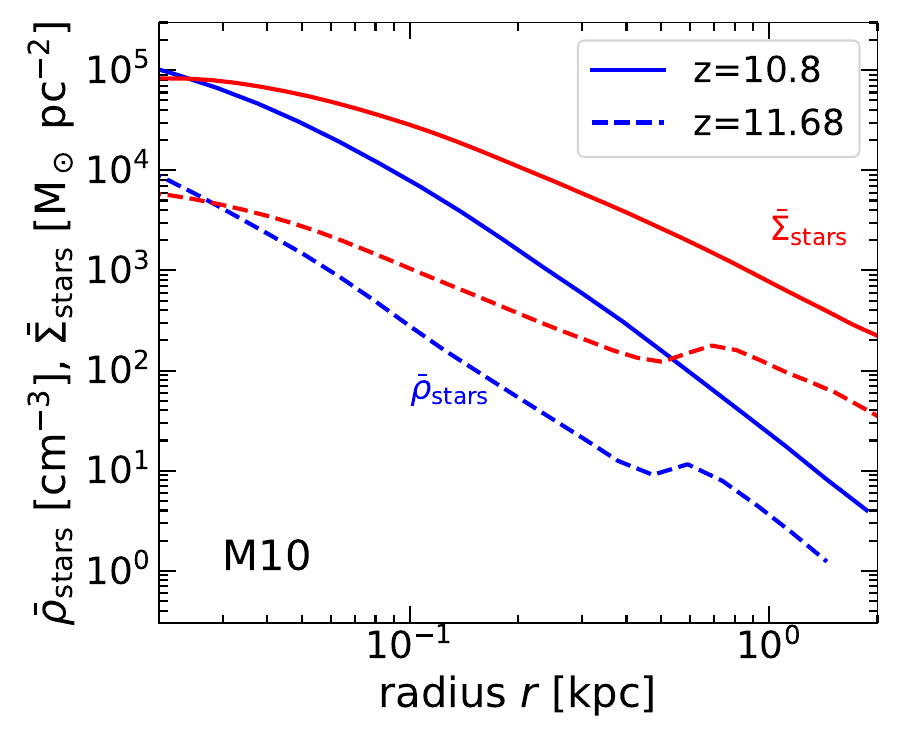}
\includegraphics[width=0.33\textwidth] {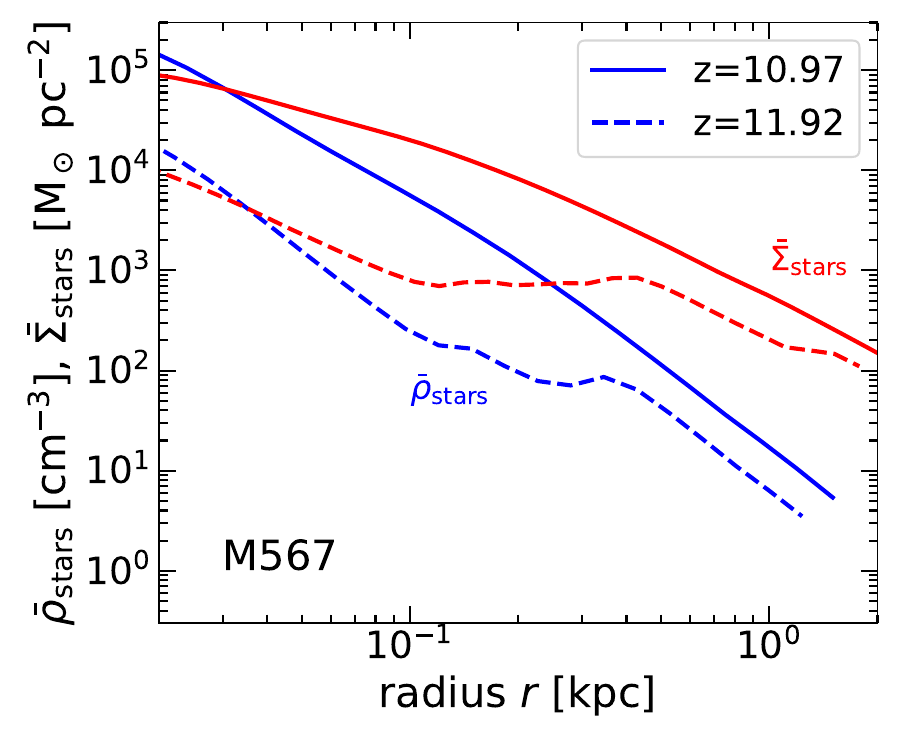}
\includegraphics[width=0.33\textwidth] {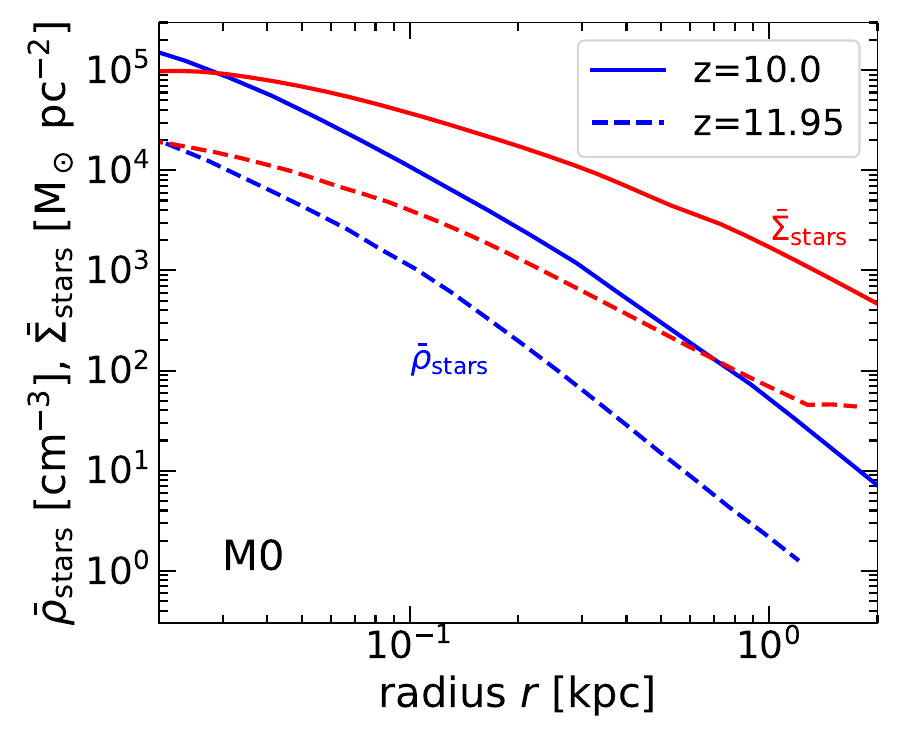}
\includegraphics[width=0.33\textwidth] {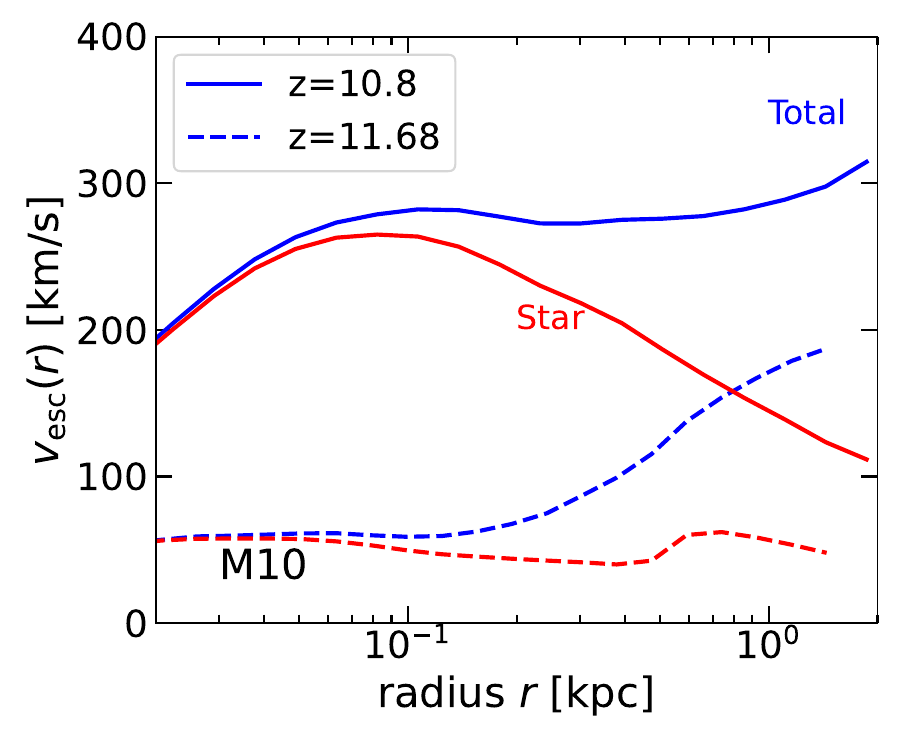}
\includegraphics[width=0.33\textwidth] {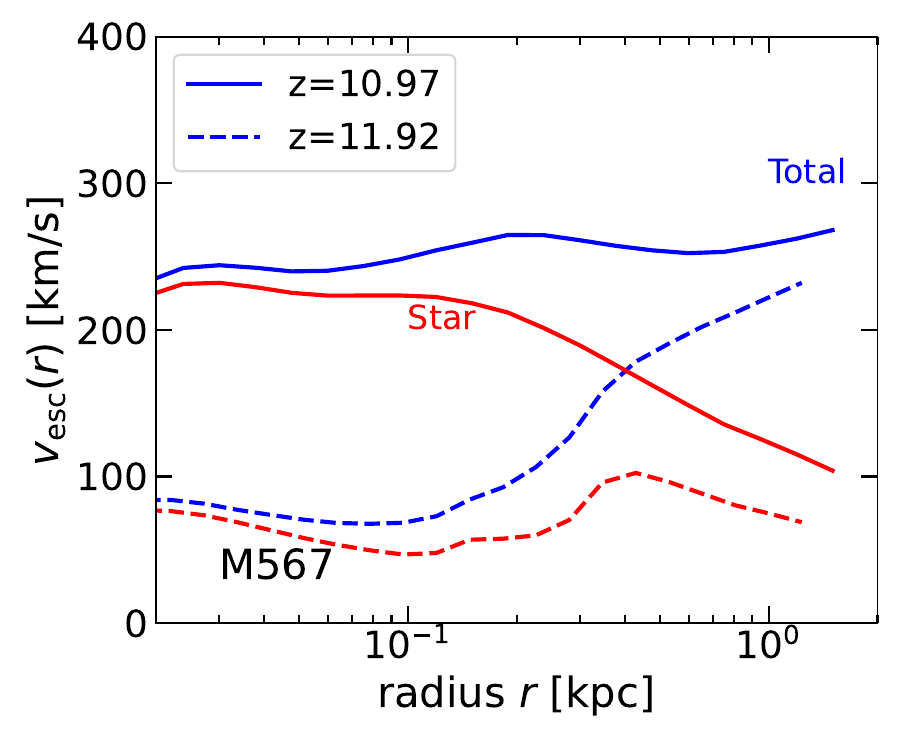}
\includegraphics[width=0.33\textwidth] {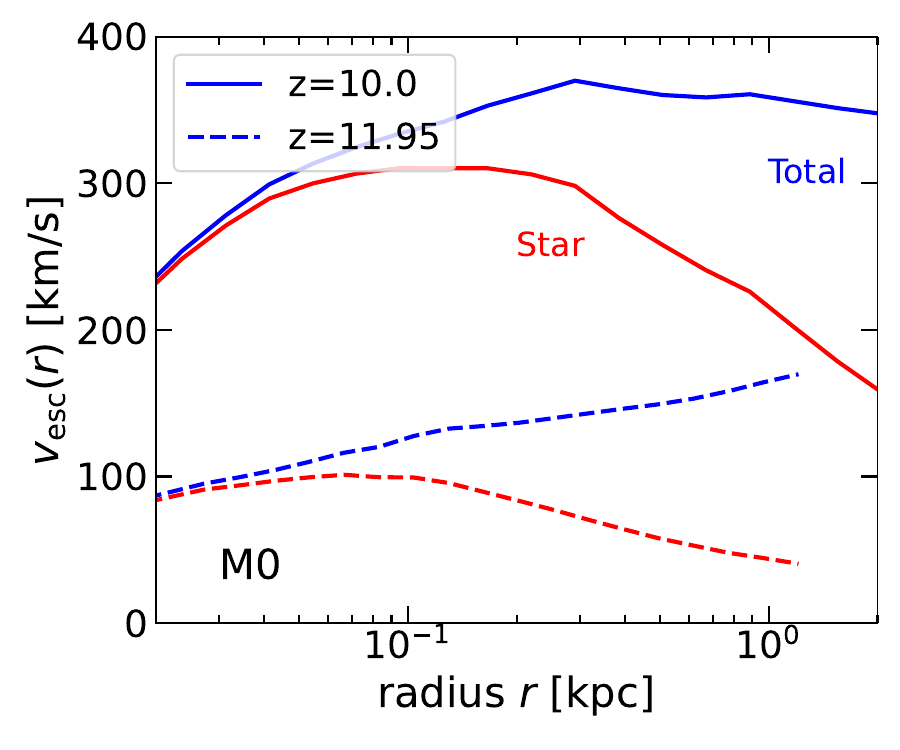}

\caption{Growth of compact central stellar clusters by wet compaction events in MAGE cosmological simulations at cosmic dawn (Yao et al., in preparation). Shown are the evolution of mass within $100\pc$ for the stars, gas and dark matter (top), the average stellar density profiles at $<r$ before and after the compaction (middle), and the escape velocity at $r$ referring to the total mass inside $r$ (bottom). The left and middle panels each refer to a major merger event. The right panel refers to multiple merger events. The growth of stellar mass within $100\pc$ during the compaction, and the corresponding densities, is by $1.43$dex, $1.30$dex, and $0.9$dex respectively. The growth of the escape velocity is by $0.70$dex, $0.62$dex, and $0.37$dex. The post-compaction stellar mass within $100\pc$ reaches $\sim\! 10^9\msun$ with densities $\bar{\rho} \seq 7 \stimes 10^3\cmc$, $5\stimes 10^3\cmc$, and $10^4\cmc$, and $\bar{\Sigma} \seq 3\stimes 10^4\msun\pc^{-2}$, $2\stimes 10^4\msun\pc^{-2}$, and $3\stimes 10^4\msun\pc^{-2}$. The escape velocities at $100\pc$ become $280\kms$, $250\kms$, and $340\kms$.}   
\label{fig:mage_prof}
\end{figure*}

%========================
\subsection{Wet Compaction in VELA Simulations at Cosmic Morning}

% VELA simulations at cosmic noon.
Wet compaction events in the VELA simulations and their major role in transforming galaxy properties have been studied in a series of papers \citep{db14, zolotov15, tacchella16_prof, tacchella16_ms, tomassetti16, dekel20_ring, dekel21_core} %, lapiner21} 
as summarized in \citet{lapiner23}.
These studies are based on the VELA suite of 32 zoom-in cosmological simulations \citep{ceverino14} using the Adaptive Refinement Tree (ART) code \citep{krav97,ceverino09}, with a maximum gas resolution of $\sim\!25\pc$. Beyond gravity and hydrodynamics, the code incorporates the physics of gas and metal cooling, UV-background photoionization, stochastic star formation, gas recycling, stellar winds and metal enrichment. The main feedback processes are thermal supernova feedback and radiative pressure from massive stars, but no
AGN feedback is included. Other limitations of these simulations are discussed in section 2 of \citet{lapiner23}. The simulated galaxy properties are summarized in Table A1 of \citet{lapiner23}. The halo masses at $z\seq 2$ range from $10^{11}\msun$ to $10^{12}\msun$. Major compaction events occur in most of these galaxies at least once during cosmic morning, in the redshift range $z \seq 2\sdash 6$ \citep[][Table A2]{lapiner23}.

\smallskip % compaction, extended disk
\Fig{vela_compaction} shows face-on images of gas and stellar densities during the stages of a typical compaction event (in galaxy V07) 
at $z \ssim 3.2$. It demonstrates the merger-driven gas contraction to a star-bursting ``blue nugget", the following passive stellar evolution to a compact ``red nugget", and the subsequent formation of an extended star-forming gas disk and ring. This disk/ring, which is fed by incoming cold streams, is stabilized by the compaction-driven formation of a central mass \citep{dekel20_ring}. The red nugget may resemble an LRD, while the appearance of an extended blue disk marks the end of the phase that is observationally identified as an LRD.

\smallskip % nuclear disk
\Fig{vela_nuclear_disk} shows zoom-in images of post-compaction densities of another VELA galaxy (V12), at $z \seq 4$, demonstrating the formation of a star-forming nuclear disk. The mean surface density within $100\pc$ is $\sim\! 5\stimes 10^5\msun\pc^{-2}$. This implies a mass of $3\stimes 10^{9}\msun$, and an escape velocity of $\sim\! 520\kms$.   

\smallskip % redshift dependence
At higher redshifts, for a given mass, the radius should scale as $(1+z)^{-1}$, and the 3D and 2D densities should scale as $(1+z)^{3}$ and $(1+z)^{2}$. As discussed below in \se{epoch}, this naturally predicts a redshift dependence of the stellar-system compactness, and therefore of the abundance of LRDs as selected by their compactness, a dependence that is consistent with the observed trend \citep{jones25}. 

\smallskip % vela compaction profiles
\Fig{vela_prof} quantifies the growth of compact central stellar systems by wet compaction events in VELA cosmological simulations at cosmic morning \citep{ceverino14, lapiner23}. We analyze two example galaxies, V07 and V12, which show the most dramatic compaction events (at $z \seq 5 \sdash 3$) among the 32 galaxies in the suite. We also analyze together the major compactions in the 16 most massive galaxies, selected by reaching a stellar mass above the median of $10^{10.06}\msun$ at $z \seq 2$. These galaxies are stacked together via time shifts that match the expansion factor $a(t)$ for each galaxy with the expansion factor $a_{\rm bn}$ that corresponds to the peak of the major compaction event in terms of gas mass within $100\pc$ (the blue nugget phase). Shown are the medians and $\pm 1\sigma$ scatter.

\smallskip
The top panels show the evolution of mass within $100\pc$ for the stars, gas, and dark matter. While \citet{lapiner23} focused on the gas mass within $1\kpc$, we here focus on the dramatic growth of inner stellar mass within $r \slt 100\pc$, relevant for LRDs. The middle panels show the average stellar density profiles $\bar{\rho}(<\!r)$ and $\bar{\Sigma}(<\!r)$ before and after the compaction. The bottom panels show the escape velocity at $r$, $\Vesc \seq \sqrt{2} V_{\rm circ}$, referring to the total mass of stars, gas and dark matter inside a sphere of radius $r$, which is relevant for the ability to retain the central BH against GW recoils. The selected pre-compaction and post-compaction times are marked in the top panels. For the stacked profiles, these times are taken to be at $\log(a/a_{\rm bn})=-0.22$ and $+0.06$.

\smallskip % results vela
We read from \Fig{vela_prof} that the growth of stellar mass and densities within $100\pc$ during the compaction is by $2.9$dex and $3.7$dex for V07 and V12, and by $2.8$dex for the stacked median. The growth of the escape velocity at $100\pc$ is by $1.4$dex, $1.4$dex, and $1.14$dex, respectively. The post-compaction stellar mass within $100\pc$ reaches values $\sim\! 3\stimes 10^9\msun$. The 3D densities are $\bar{\rho} \seq 4 \stimes 10^4\cmc$, $4\stimes 10^4\cmc$, and $10^4\cmc$ for the stacked median, and the surface densities are $\bar{\Sigma} \seq 5\stimes 10^4\msun\pc^{-2}$, $5\stimes 10^4\msun\pc^{-2}$, and $2\stimes 10^4\msun\pc^{-2}$ for the median. The escape velocities at $100\pc$ become $540\kms$, $540\kms$, and $405\kms$ for the stacked median, with the $+1\sigma$ value reaching $560\kms$.

\smallskip
We learn that the growth of stellar mass and densities within $100\pc$ due to the compaction event alone can be extremely large, by up to three orders of magnitude. This leads to central densities comparable to those of LRDs, and to escape velocities that are at the level required for retaining the SMBH against GW recoils. We assume that adding the effect of dry mergers of clusters could make the growth even more dramatic.

%=================
\subsection{Wet Compaction in other simulations at Cosmic Morning}

It is worth referring to other simulations that explore wet compaction events at cosmic morning. We refer in particular to the analysis of FirstLight simulations by \citet{cataldi25}, and to the direct address of how a compaction leads to an LRD-like system by \citet{mayer24}.  

\smallskip % cataldi
\citet{cataldi25} used 169 galaxies from the FirstLight suite of cosmological simulations \citep{ceverino17} to study wet compaction events during cosmic morning, $z \seq 5 \sdash 9$ (compared to 32 galaxies with compaction events at $z \seq 1 \sdash 7$ in the VELA simulations). These simulations used the same ART code \citep{ceverino09} with a somewhat higher maximum resolution for the gas of $8.7 \sdash 17$ proper parsecs (compared to $\sim\!25\pc$ in the VELA simulations). They focused on identifying compaction events based on the evolution pattern of the stellar half-mass radius (compared to evolution of gas and stellar mass in VELA). Their results largely strengthen the results from the VELA simulations \citep{lapiner23}. They find the favorable stellar mass range for compaction to be $\Ms \seq 10^{8} \sdash 10^{10}\msun$ \citep[][Fig.~4]{cataldi25}, similar to the range found in VELA \citep[][Fig.~10]{lapiner23}. They find that the half-mass radius typically shrinks by $0.5 \sdash 1.0$dex during the compaction, sometimes down to $100\pc$ \citep[][Fig.~3]{cataldi25}, very similar to the finding in VELA \citep[][Fig.~14]{lapiner23}. Finally, they recover in a significant fraction of the galaxies a post-compaction increase in the half-mass radius, associated with the growth of an extended disk as seen in VELA, which we propose to identify with the end of the LRD phase (\se{epoch}).

\smallskip % Mayer
Appealing more directly to the formation of LRDs at cosmic morning, \citet{mayer24} studied  the formation of a compact stellar system with a massive BH by a merger-driven compaction event in a cosmological simulation. Using the SPH code GASOLINE2 \citep{wadsley17}, with softening $\sim\!100\pc$ and maximum smoothing length of $7\pc$, they analyze a major-merger remnant at $z \ssimeq 8$. They find that about 10\% of the gas in the host galaxy is compactified into a supermassive nuclear disk (SMD) of $\sim\!10\pc$ in size, a nuclear starburst and $3 \stimes 10^8\msun$ in gas. Being bar unstable \citep{mayer15, mayer19, zwick23}, they expect further contraction, which will become general-relativistic-unstable and form a supermassive BH of $10^6 \sdash 10^8\msun$ \citep{shibata02, saijo09, zwick23}. In \citet{zwick25}, they modeled the spectral features of an SMD of this sort, and showed that they match the spectra of LRDs, where the V-shape arises from the superposition of two black bodies, and the Balmer line broadening is sourced by the intrinsic rotation of the SMD. This model may work without an AGN and with no X-rays. They argue that the number densities of LRDs are reproduced with mergers of galaxies more massive than $10^8\msun$.

%=====================
\subsection{Wet Compaction in MAGE Simulations at Cosmic Dawn}

% yao simulations
In order to further explore the effects of wet compactions, with a different code and physical recipes and at earlier epochs corresponding to cosmic dawn, we analyze the compaction events in three galaxies from the MAGE suite of eleven zoom-in cosmological simulations (Yao et al., in preparation). This suite is utilizing the Adaptive Mesh Refinement (AMR) RAMSES code \citep{teyssier02} with $10\pc$ resolution for the gas. The physical models used are described in \citet{andalman25}. The galaxies M10 and M567 each happen to have a major merger (mass ratio $\sim\!1$ and $\sim\!0.5$, respectively) at $z \ssim 12\sdash 11$. The galaxy M0 has three successive mergers (of mass ratios $\sim\! 0.5, 0.6, 0.3$) at $z \ssim 11.5\sdash 10$. Two more galaxies from the suite of eleven (not shown here) have major mergers in the redshift range $13 \sdash 9$.

\smallskip %mage compaction profiles
\Fig{mage_prof} quantifies the growth of compact central stellar systems by wet compaction events in the three MAGE simulations, along the lines of the analysis of VELA compaction events shown in \Fig{vela_prof}. The focus is on the mass, densities and escape velocity within $100\pc$, relevant for LRDs.

\smallskip % results mage
We read from \Fig{mage_prof} that the growth of stellar mass and densities within $100\pc$ during the compaction is by $1.4$ dex, $1.3$ dex, and $0.9$ dex respectively. The growth of the escape velocity is by $0.7$ dex, $0.6$ dex, and $0.4$ dex. The post-compaction stellar mass within $100\pc$ reaches $\sim\! 10^9\msun$ with a 3D density $\bar{\rho} \seq 7 \stimes 10^3\cmc$, $5\stimes 10^3\cmc$, and $10^4\cmc$ for the three galaxies, and a surface density $\bar{\Sigma} \seq 3\stimes 10^4\msun\pc^{-2}$, $2\stimes 10^4\msun\pc^{-2}$, and $3\stimes 10^4\msun\pc^{-2}$, respectively. The escape velocities at $100\pc$ become $280\kms$, $250\kms$, and $340\kms$.
 
\smallskip
We learn that the growth of stellar mass and densities within $100\pc$ due to the compaction event alone can be $(1 \sdash 1.5)$ dex. This leads to central densities comparable to those of LRDs. The escape velocities are not by themselves at the level required for retaining the SMBHs against GW recoils. The lower escape velocities in MAGE compared to VELA is partly due to the fact that the MAGE galaxies at cosmic dawn are less massive than the VELA galaxies at cosmic morning. However, the growth of escape velocity by $(0.4 \sdash 0.7)$ dex in the MAGE galaxies does provide a significant deepening of the potential well. We assume that once combined with the growth induced by dry mergers of clusters it can bring the escape velocity to the level required for retaining the SMBHs against GW recoils.

%%%%%%%%%%%%%%%%%%%%%%%%%%%% 5
\section{Epoch, Mass and Abundance of LRDs}
\label{sec:epoch}

If LRDs form along the lines of the scenario outlined above, one wishes to understand how it determines the epoch in which they are observed at cosmic morning, namely $z \simeq 4 \sdash 8$, and their characteristic stellar masses of $\lsim\! 10^{9}\msun$.

\smallskip % appearance
The appearance of LRDs at $z \ssim 8$ stems from the crucial need for star clusters in the proposed scenario. They are needed first of all for generating the BH seeds, and then for the efficient migration into the galaxy center where they can merge into a compact stellar system and allow the BH seeds to merge. These clusters naturally form in the FFB phase, typically prior to $z \simeq 8$ \citep[][Fig.~6]{dekel23}, and can thus lead to LRD formation soon after. The galactic stellar mass for a fiducial FFB galaxy is predicted to be $\Ms \ssim 2\stimes 10^9\msun\, \epsilon_{0.2}\, (1+\zffb)_{10}^{-6.2}$, where $\epsilon \sequiv 0.2\,\epsilon_{0.2}$ is the global integrated star-formation efficiency, and $(1+\zffb) \sequiv 10\,(1+\zffb)_{10}$ refers to the redshift of the FFB phase. The total stellar mass in the clusters is thus predicted to be of order $\sim\!10^9\msun$, and smaller if the FFB phase occurred at $z \sgt 9$ \citep[][Fig.~6]{dekel23}.

\smallskip %- Why do LRD stop at $z<4$ and above $10^9$?    
The apparent disappearance of LRDs from the observational samples towards cosmic noon and above a certain stellar mass may actually be a result of the same compaction events that are 
the second key element for LRD formation according to the proposed scenario. This is the post-compaction generic formation of an extended gaseous, star-forming disk or ring, as seen in \Fig{vela_compaction} and discussed in \citet{lapiner23}. The shrinkage of this disk/ring is suppressed by the compaction-driven formation of a massive central object 
\citep{dekel20_ring}. This adds an extended blue envelope around the red LRD, which makes it identified as non-LRD. It is ironic that the same compaction events that allow the appearance of LRDs with BHs eventually lead to their disappearance from the observed samples.

\smallskip % elbaz
Observational support for the disappearance of LRDs by the formation of a blue envelope is provided by the detection of a population of galaxies with compact LRD-like central regions and star-forming blue envelopes at $z \slt 4$, using CEERS/JWST images \citep{billand25}. The properties of these galaxies and their abundance seem to be consistent with being the post-LRD descendants of the LRDs. We note that, while this analysis assumed the stars-only model for LRDs and post-LRDs, namely very high stellar masses of $10^{10}\msun$ and surface densities of $10^7\msun\pc^{-2}$, the match of these populations is not limited to this model. Similar results can be obtained when adopting the hybrid model of stars plus BHs both for the LRDs and the post-LRDs (David Elbaz, private communication).

\smallskip % z=4 Ms=9
Why would this scenario cause disappearance of LRDs after $z \ssim 4$ and above stellar masses of $\sim\! 10^9\msun$ (as deduced for LRDs with BHs)? It has been shown using analytic modeling and cosmological simulations that extended disks can be long-lived against spin-flips due to further major mergers once the halo mass is above $\sim\! 10^{11}\msun$ 
\citep{dekel20_flip}. For the typical SFE of $\sim\!0.1$ at late cosmic morning, this implies a stellar mass above $\sim\! 10^9\msun$. These halos represent two-sigma peaks in the density fluctuation field at $z \ssim 4$, implying that more massive halos that can accommodate long-lived extended disks become more common after $z \ssim 4$.

\smallskip %fasifiable predictions
Using the VELA simulations, properties of extended, long-lived rings have been studied in \citet{dekel20_flip}. We refer the reader to Figure B1, which can be used to make predictions about post-LRD descendants in our proposed LRD disappearance scenario. We learn that the median ring radius is $0.5 R_{\rm d}$, where $R_{\rm d}$ is the galaxy half mass radius, and that the median specific star formation rate in the ring is $\sim 1\ \rm Gyr^{-1}$, about $\sim 1$ dex higher than that in the core interior to it. For $10^{11}\msun$ halos at $z \sim 4$, $R_{\rm d} \sim 500\ {\rm pc}$, assuming $\lambda = 0.025$, and the median ring radius turns out to be $\sim 250\pc$. While the VELA galaxies being at $z \sim 1.5$ are more massive with median stellar masses of $\sim 10^{10} M_{\odot}$, these results may still qualitatively hold for post-compaction rings formed in $\sim\! 10^9\msun$ galaxies at $z \sim 4$. Since we do not model SEDs, if LRDs are defined as compact systems with effective sizes less than $300 \pc$ \citep[][]{pacucci25a}, this galaxy with a compact core + extended ring would have moved out of the LRD sample. While we don't explicitly model it, the blue, star forming envelope would also make the characteristic V-shaped LRD SED less distinct, generally moving the galaxy out of the LRD sample in color-magnitude or color-color selection diagrams as well.

\smallskip % abundance
The number density of LRDs with BHs at cosmic morning is predicted to roughly follow the number density of FFB galaxies at cosmic dawn, which is predicted to be $n \ssim 10^{-5} \sdash 10^{-4} \Mpc^{-3}$ \citep{dekel25_post}. This stems from the prediction that cluster-dominated galaxies are a generic feature of the FFB phase, and because, as seen in the VELA simulations, every galaxy is expected to undergo at least one major compaction event during the $\sim\!1\Gyr$ period from cosmic dawn to cosmic morning. This predicted number density is in the ball park of the observed abundance of LRDs \citep{kokorev24, greene24, kocevski25}. However, beyond this crude estimate, we acknowledge that the formation of LRDs via our proposed channel requires several different processes to work synergistically, namely, efficient seed BH formation in each FFB cluster, efficient migration of the clusters to the central region (depends on disk and cluster properties), efficient BH mergers overcoming the GW recoil bottleneck (depends on  merging BH mass ratio, spins, and spin-orbit misalignment), which can be aided by a wet-compaction event if it occurs relatively soon after the FFB phase, as the clusters will migrate in on timescales of $\sim 100\ {\rm Myr}$. While the results from \citet[][]{chen26}, who resolve the formation of FFB clusters in a zoom-in cosmological simulation and also find efficient cluster migration in the inner galaxy are encouraging, robust LRD abundances via this formation channel can only be estimated from a suite of cosmological zoom-ins that self-consistently simulate the formation of a disk of FFB clusters along with their seed BHs, the subsequent dry migration of the clusters, wet compaction, and growth of the central supermassive BH, starting from different initial conditions (halo properties and cosmic environment). This is a challenging task that requires sub-pc resolution and may even then be only possible with some sub-grid recipe for seed black hole formation.

\smallskip % redshift dependence
The fact that at a given mass the expected radii scale with $(1+z)^{-1}$ naturally lead to higher compactness at higher redshifts, and thus to an abundance of detected LRDs that is growing with redshift during cosmic morning, as indicated by observations \citep{jones25}.

%%%%%%%%%%%%%%%%%%%%%%%%%%%%%%%%%%%%% 
\section{Conclusion}
\label{sec:conc}

We propose that the Little Red Dots, as detected in JWST observations, form naturally at cosmic morning, $z \seq 4 \sdash 8$, as compact stellar systems with over-massive black holes. The two key elements in the proposed scenario are (a) the thousands of star clusters that form at cosmic dawn, $z \sgt 8$, in a phase of feedback-free starbursts within the clusters, and (b) wet compaction events that are associated with mergers or colliding cold streams. Our conclusions regarding four central elements of LRD formation can be summarized as follows.

\subsection{Dry evolution of clusters}

Investigating the dry evolution of the galactic disk of FFB clusters using analytic toy-modeling, we learned the following:

\bul
The $10^6\msun$ clusters allow efficient migration to the galaxy center, mostly by two-body segregation aided by dynamical friction against the disk, on a timescale of $\sim\!100\Myr$ (and shorter for more massive clusters).

\bul
The clusters merge into compact central stellar systems, containing $M \ssim 10^9\msun$ within radii of $R \ssim 60\pc$, These are already in the ball park of the stellar systems deduced from observed LRDs using a hybrid model of stars and BHs.

\bul
Mutual tidal stripping leaves most of the cluster mass intact such that it does not make a qualitative difference to the analysis. The part of the mass that is stripped into a smooth component allows the disk to assist the migration by exerting dynamical friction on the clusters.  

\bul
Simulations of dry evolution of clusters with BHs (Dutta Chowdhury et al., in preparation) and high-resolution cosmological simulations at cosmic dawn \citep[][]{chen26}, reproduce the predicted rate of cluster migration and the formation of a massive, compact, central stellar system via cluster-cluster mergers.

\subsection{Over-massive black holes}

Regarding the growth of over-massive black holes in the LRDs, we find the following:

\bul
\citet{dekel25_bh}, following others, demonstrated that the FFB clusters are natural sites for the formation of BH seeds via rapid core collapse. The core collapse into a $10^4\msun$ BH seed occurs in less than $3\Myr$ in young, rotating clusters of $\sim\!10^6\msun$. It is driven by gravo-thermal instability, which is governed by the massive stars that are still present during the first $\sim\!3\Myr$. The core collapse is sped by gravo-gyro instability as long as the clusters are supported by rotation induced by their host disks. 

\bul
The migrating clusters then carry with them the BH seeds, making them ready for merging at the galaxy center into an over-massive BH.

\bul
Extending the analysis of \cite{dekel25_bh}, we showed that further deepening of the central potential well is needed in order to retain the SMBH in the galaxy against post-merger
GW recoils, and in order to lock it to the galaxy center. This is higher than the compactness produced by dry migration of clusters, which is expected to lead to an escape velocity of $\Vesc \ssim 300\kms$. 

\subsection{Wet compaction events}

We proposed that a natural solution to the GW recoil problem is provided by the common wet compaction events, triggered by mergers or other gaseous mechanisms that involve drastic energy and angular-momentum losses \citep{lapiner23}. In order to analyze the compaction-driven growth of a massive, compact, central stellar system, we used a variety of cosmological simulations, based on different codes and physical recipes, with compaction events that occur at redshifts from cosmic dawn to cosmic noon. We find the following:

\bul
The relative increase of stellar mass within $100\pc$ in major compaction events can range from $0.5$dex to more than $3$dex. This by itself can lead to central stellar systems 
that match or exceed the compactness of LRDs as deduced from observations using the hybrid model of stars and BHs \citep[e.g.,][]{pacucci24a,pacucci24b}.

\bul
Major compactions typically boost the escape velocities by $(0.5 \sdash 1.5)$dex. Adding to the growth of the central stellar system by dry migration and mergers of clusters, we assume that the compactions can increase the escape velocities to the level required for retaining the SMBHs against GW recoils. Alternatively, these two mechanisms together could produce central stellar densities as high as required for LRDs without BHs. 

\subsection{Epoch, masses and abundance of LRDs}

The period when LRDs are typically detected, $z \seq 8$ to $4$, and their favorable stellar masses of $\lsim\! 10^9\msun$, can be related to the key elements of the scenario, 
namely clusters and compaction events, as follows:

\bul
The LRDs appear after the formation of FFB clusters at cosmic dawn, typically at $z \ssim 8$ \citep{dekel23,dekel25_post}, as they are crucial for the LRD formation. The fiducial stellar masses of FFB galaxies are predicted to be $\sim\! 10^9\msun$.

\bul
The LRDs escape being classified as LRDs by the natural development of post-compaction blue extended disks around the ``LRD" cores. These disks/rings are fed by cold streams and are stabilized against contraction by the presence of compaction-driven central masses \citep{dekel20_ring}. The disks tend to survive subsequent mergers once the disk mass is above a threshold of $\sim\! 10^9\msun$ \citep{dekel20_flip}. At $z \ssim 4$, the dark-matter halos of corresponding masses become common two-sigma peaks of the density fluctuation field.

\bul
The number density of LRDs with BHs at cosmic morning is predicted to follow the number density of FFB galaxies, $n \ssim 10^{-5} \sdash 10^{-4} \Mpc^{-3}$ \citep{dekel25_post}, in the ball park of the observed abundance of LRDs. This is because cluster-dominated galaxies are a generic feature of the FFB phase and because every galaxy is expected to undergo major compaction events. 

\bul
The abundance of LRDs is expected to increase from $z \ssim 4$ to $z \ssim 8$ due to the scaling of size with $(1+z)^{-1}$ at a given mass, as observed.
\citep{jones25}.

\smallskip % future simulations
The exploration of LRD formation by the combined effects of cluster migration and global compaction events will require challenging cosmological simulations with sub-parsec resolution that can capture the FFB physical processes in the clusters. They will have to rely on BH seed formation in the clusters, and simulate the migration and mergers of the clusters with BHs into LRD-like compact stellar systems with SMBH

\smallskip %alternate
A alternative formation channel identifies LRDs with quasi-stars, namely rapidly accreting direct collapse black holes \citep[][]{natarajan24} embedded in massive, optically thick, radiation-supported gaseous envelopes \citep[][]{begelman08, degraff25, begelman26, naidu26, nandal26, santarelli26}. \citet{santarelli26} report that the quasi-star phase is short-lived lasting for $\sim 20$--$40\ {\rm Myr}$ before becoming a $10^6 M_{\odot}$ SMBH. In contrast, typical timescales for $\gtrsim 10^{6} M_{\odot}$ SMBH formation by migration and dry mergers of $\lesssim 10^{6} M_{\odot}$ FFB clusters, carrying $\lesssim 10^4 M_{\odot}$ seed black holes is $\sim 100 \Myr$. In our proposed scenario, the compact stellar component of the LRD assembles in parallel to the SMBH assembly, in fact the SMBH assembly is aided by the former, while in the quasi-star model, an additional host galaxy build up is required.

%%%%%%%%%%%%
\section*{Acknowledgments}

This paper is dedicated to the memory of Avishai Dekel and Joel Primack, who are no longer with us. Joel passed away shortly before its submission, and Avishai shortly afterward, within days of one another. We are deeply grateful to have known and worked with them and to have benefited from their mentorship and support. Their dedication to astrophysics and passion for scientific inquiry will continue to inspire us. We thank the anonymous referee for their insightful comments that led to improvements in the manuscript. We are grateful for stimulating discussions with Hou-Zun Chen and Zhaozhou Li. This work was supported by the Israel Science Foundation Grant ISF 861/20, by the US National Science Foundation - US-Israel Bi-national Science Foundation grants 2023730 and 2023723, and in part by grant NSF PHY-2309135 to the Kavli Institute for Theoretical Physics (KITP). DDC acknowledges support from the Excellence Fellowship Program for International Postdoctoral Researchers by the Council for Higher Education and the Israel Academy of Sciences and Humanities.

%%%%%%%%%%%%%%%%%%

\bibliography{z10}{}
\bibliographystyle{aasjournalv7}

\end{document}